%% file: autosam-LPV-SS.tex
\documentclass[twocolumn]{autart}    
                               
\usepackage[english]{babel}															
\usepackage{graphics} 
\usepackage{mathtools} 
\usepackage{times} 
\usepackage{amsmath} 
\usepackage{amssymb}  
\usepackage{amsfonts}%
\usepackage{euscript}
\usepackage{mathrsfs}
\usepackage[inline]{enumitem}
\usepackage{xcolor}
\usepackage{multirow,tabularx}
\usepackage{subcaption}
\usepackage{hhline}
\usepackage{algcompatible}
\usepackage{algorithm}
\usepackage{cite}
\usepackage{eufrak}
\usepackage[normalem]{ulem} 
\usepackage[
  bookmarks=false,
  pdfpagelabels=false,
  hyperfootnotes=false,
  hyperindex=false,
  pageanchor=false,
  implicit=false,
  colorlinks=false,
]{hyperref}
\usepackage{tikz}

\usetikzlibrary{shadows.blur,positioning,backgrounds,fit} 
\usepackage{pgfplots}
\pgfplotsset{compat=newest} 
\pgfplotsset{plot coordinates/math parser=false}

\DeclareMathOperator*{\argmin}{\arg\!\min~}
\DeclareMathOperator*{\argmax}{\arg\!\max~}

\newcommand{\NX}{{n_\mathrm{x}}}

\newcommand{\NY}{{n_\mathrm{y}}}
\newcommand{\NU}{{n_\mathrm{u}}}
\newcommand{\NP}{{n_\mathrm{p}}}
\newcommand{\NPSI}{{n_{\psi}}}			
\newcommand{\NO}{{n_\mathrm{o}}}
\newcommand{\NR}{{n_\mathrm{r}}}
\newcommand{\NH}{{n_\mathrm{h}}}
\newcommand{\NF}{{n_\mathrm{f}}}
\newcommand{\NTH}{{n_\mathrm{\theta}}}

\newcommand{\sU}{\mathbb{U}}
\newcommand{\sP}{\mathbb{P}}
\newcommand{\sX}{\mathbb{X}}
\newcommand{\sY}{\mathbb{Y}}
\newcommand{\sZ}{\mathbb{Z}}
\newcommand{\sI}{\mathbb{I}}

\newcommand{\Ru}{\mathbb{R}^{n_\mathrm{u}}}
\newcommand{\Rp}{\mathbb{R}^{n_\mathrm{p}}}
\newcommand{\Rx}{\mathbb{R}^{n_\mathrm{x}}}
\newcommand{\Ry}{\mathbb{R}^{n_\mathrm{y}}}

\newcommand{\Ym}{{\overline{Y}}}
\newcommand{\Phim}{{\overline{\Phi}}}
\newcommand{\thetam}{{\overline{\theta}}}
\newcommand{\Vm}{{\overline{W}}}

\newcommand{\ind}[1]{{[#1]}}

\newcommand{\sys}{\mathcal{S}}

\newcommand{\M}{\EuScript{M}}

\newcommand{\Qv}{\EuScript{Q}}
\newcommand{\Sv}{\EuScript{S}}
\newcommand{\Rv}{\EuScript{R}}
\newcommand{\Vv}{\EuScript{V}}

\newcommand{\reach}{\mathcal{R}}
\newcommand{\obsv}{\mathcal{O}}
\newcommand{\hank}{\mathcal{H}}

\newcommand{\corr}{\mbox{R}}
\newcommand{\expct}{\mathbb{E}}

\newcommand{\selO}{\nu}
\newcommand{\selR}{\varsigma}

\newcommand{\rank}{\mathrm{rank}}

\newcommand{\var}{\mathrm{var}}

\newcommand{\SNR}{\mathrm{SNR}}
\newcommand{\MSE}{\mathrm{MSE}}
\newcommand{\BFR}{\mathrm{BFR}}

\newcommand{\eye}{{\mathrm{I}}}
\newcommand{\vecM}{{\mathrm{vec}}}

\newcommand{\RWLS}{{\mathrm{RWLS}}}

\newcommand{\Afnc}{\mathcal{A}}
\newcommand{\Bfnc}{\mathcal{B}}
\newcommand{\Cfnc}{\mathcal{C}}
\newcommand{\Dfnc}{\mathcal{D}}
\newcommand{\Gfnc}{\mathcal{G}}
\newcommand{\Hfnc}{\mathcal{H}}
\newcommand{\Kfnc}{\mathcal{K}}
\newcommand{\Pfnc}{\mathcal{P}}

\newcommand{\uniform}{\mathcal{U}}
\newcommand{\Dat}{\EuScript{D}_N}
\newcommand{\Dval}{\EuScript{D}_\mathrm{val}}

\newcommand{\numstr}[1]{\#(#1)}

\newcounter{ProbInt}   
\refstepcounter{ProbInt}

\newcommand{\probint}{Problem \arabic{ProbInt}\refstepcounter{ProbInt}}

\protected\edef\ell{\noexpand\ensuremath{{\mathchar\the\ell}}} 

\begin{document}

\begin{frontmatter}

\title{Towards Efficient Maximum Likelihood \\ Estimation of LPV-SS Models\thanksref{footnoteinfo}\thanksref{footnoteinfo2}} 

\thanks[footnoteinfo]{This paper was not presented at any IFAC 
meeting. Corresponding author P.~B.~Cox. Tel. +31-40-2478188.}
\thanks[footnoteinfo2]{This paper has received funding from the European Research Council (ERC) under the European Union's Horizon 2020 research and innovation programme (grant agreement No 714663).}

\author[CSTUE]{Pepijn B. Cox}\ead{p.b.cox@tue.nl},    
\author[CSTUE]{Roland T\'{o}th}\ead{r.toth@tue.nl},               
\author[Lille]{Mih\'{a}ly Petreczky}\ead{mihaly.petreczky@ec-lille.fr}

\address[CSTUE]{Control Systems Group, Department of Electrical Engineering, Eindhoven University of Technology, P.O. Box 513, \\ 5600 MB Eindhoven, The Netherlands.}  
\address[Lille]{Univ. Lille, CNRS, Centrale Lille, UMR 9189 - CRIStAL - Centre de Recherche en Informatique Signal et Automatique de Lille,\\ F-59000 Lille, France.}   


\begin{abstract}                          
How to efficiently identify multiple-input multiple-output (MIMO) linear parameter-varying (LPV) discrete-time state-space (SS) models with affine dependence on the scheduling variable still remains an open question, as identification methods proposed in the literature suffer heavily from the curse of dimensionality and/or depend on over-restrictive approximations of the measured signal behaviors. However, obtaining an SS model of the targeted system is crucial for many LPV control synthesis methods, as these synthesis tools are almost exclusively formulated for the aforementioned representation of the system dynamics.  Therefore, in this paper, we tackle the problem by combining state-of-the-art LPV input-output (IO) identification methods with an LPV-IO to LPV-SS realization scheme and a maximum likelihood refinement step. The resulting modular LPV-SS identification approach achieves statical efficiency with a relatively low computational load. The method contains the following three steps: 
\begin{enumerate*}[label=\arabic*)] 
	\item estimation of the Markov coefficient sequence of the underlying system using correlation analysis or Bayesian impulse response estimation, then
	\item LPV-SS realization of the estimated coefficients by using a basis reduced Ho-Kalman method, and
	\item refinement of the LPV-SS model estimate from a maximum-likelihood point of view by a gradient-based or an expectation-maximization optimization methodology. 
\end{enumerate*}
The effectiveness of the full identification scheme is demonstrated by a Monte Carlo study where our proposed method is compared to existing schemes for identifying a MIMO LPV system.
\end{abstract}
\vspace{-4mm}
\end{frontmatter}


\section{Introduction}
\vspace{-2mm}

The \textit{linear parameter-varying} (LPV) modeling paradigm offers an attractive model class to capture nonlinear and/or time-varying systems with a parsimonious parameterization. The LPV model class preserves the linear signal relation between the inputs and outputs of the system, however, these linear relations are functions of a measurable, time-varying signal, the scheduling variable, denoted as $p$. This scheduling signal can be any combination of inputs, measurable process states, outputs, or measurable exogenous variables and, in addition, these signals can be filtered by any arbitrary functional relation. Hence, the LPV modeling paradigm can represent both non-stationary and nonlinear behavior of a wide variety of physical or chemical processes, e.g., see~\cite{GrootWassink2005,Veenman2009,Wingerden2009,Bachnas2014}.

The majority of LPV control synthesis methods are based upon the assumption that an LPV \emph{state-space} (SS) model of the system is available, especially with static and affine dependence of the involved matrix coefficients on the scheduling variable $p$, e.g.,~\cite{Mohammadpour2012}. Hence, efficient identification of LPV-SS models in terms of computational load, statistical, and performance properties has intensively been researched. Conceptually, LPV identification can be performed as: 
\begin{enumerate*}[label=\roman*)]
	\item the interpolation of local LTI models estimated from multiple experiments around fixed operating points, i.e., with constant $p$, often referred to as the \textit{local identification} setting; or 
	\item a direct model estimation problem, i.e., the \textit{global identification} setting, which requires the experimental data with a varying $p$ which is informative to uniquely identify the considered model parameters.
\end{enumerate*}
Accordingly, global identification approaches include scheduling dynamics, see~\cite{Bachnas2014} for a detailed comparison between the two settings. In this paper, we will focus on the global setting and the identification of discrete-time models.

In the global setting, an attractive identification approach is the minimization of the $\ell_2$-loss in terms of the prediction-error associated with the model. Approaches aiming at this objective are often called \emph{prediction-error methods} (PEM). Early approaches in the PEM setting are proposed under the unrealistic assumption of full state measurements~\cite{Nemani1995,Rizvi2015}. To overcome this assumption and to directly minimize the $\ell_2$ loss, \textit{gradient-based} (GB) methodologies have been introduced, e.g., see~\cite{Lee1997a,Verdult2003,Wills2008,Wills2011}. Recently, an \textit{expectation-maximization} (EM) algorithm has been developed for LPV-SS models~\cite{Wills2011}, extending the set of GB methods. The EM method is more robust to an inaccurate initial estimate compared to the GB PEM; however, its convergence rate is much slower near the optimum~\cite{Watson1983}. Due to the nonlinear optimization associated with the EM and GB methods, their convergence to the \textit{maximum-likelihood} (ML) estimate depends heavily on a proper initial seeding. Besides prediction-error identification methods, LPV grey-box~\cite{AngelisPHD,Gaspar2007} and LPV \emph{set-membership} (SM)~\cite{Bianchi2009,Novara2011,Ceronea2013} identification approaches have been developed. Grey-box schemes require detailed knowledge of the dynamical structure of the system with only a few unknown parameters, which are often estimated by a Kalman like filtering strategy. The SM methods characterize noise and disturbances in a deterministic bounded-error compared to the stochastic description in PEM. In general, SM approaches have a significantly higher computational load compared to direct PEM and rely on convex outer-approximations. Hence, in order to achieve stochastically interpretable and computationally attractive identification of LPV-SS models, it is favorable to apply GB and EM based PEM. However, these methods require a proper initial estimate close to the global optimum (ML estimate) in order to exploit their advantageous properties (\probint).

To achieve initialization of direct PEM, alternative methods can be introduced that rely on realization theory by sacrificing ML properties for an estimation problem solvable via convex optimization. These methods boil down to: first identifying an LPV-IO model, with well-established methods available in the literature (e.g., see~\cite{Mohammadpour2012,Laurain2010,Santos2011}); and, secondly, to execute an exact realization of the identified LPV-IO form to an LPV-SS model. However, such an exact realization will, in general, result in relations with rational, dynamic dependence on the scheduling variable or lead to a non-minimal state realization if the static, affine dependence is enforced to be preserved~\cite{Toth2010a}. Moreover, such exact algebraic realization methods have a high computational cost. Recently introduced LPV realization theory based schemes, so-called \textit{subspace identification} (SID) methods, aim to avoid the aforementioned problem by achieving data-driven state-space realization. SID schemes can apply \textit{direct} LPV Ho-Kalman like realization~\cite{Toth2012} to obtain the SS matrices from specific LPV-IO models that are identified by a least-squares method; or have an intermediate \textit{projection} step, i.e.,
\begin{enumerate*}[label=\arabic*)] 
	\item identify an IO structure using convex optimization,
	\item find a projection to estimate the unknown state-sequence via matrix decomposition methods, then
	\item estimate the SS matrices in a least-squares fashion,
\end{enumerate*}
e.g., see~\cite{Felici2007a,Santos2007,Larimore2013,Wingerden2009a}. However, to attain a convex problem, the latter class of SID methods usually depend on over-restrictive approximations of the signal behaviors and/or the number of observed variables grows exponentially. As a consequence, the estimation problem has still a high computational demand, making it inapplicable for real-world systems. The aforementioned realization based schemes provide an LPV-SS model estimate which is not minimized w.r.t. any criterion and, therefore, it is not ``optimal'' in an ML sense. Hence, to solve Problem 1, i.e., to have efficient initialization of direct PEM methods, we require novel computationally attractive SS identification methods capable of providing estimates that are sufficiently close to the global PEM optimum. \probint: Finding initial estimates in the region of attraction to the ML solution in a computationally attractive way.

Based on Problems 1 and 2, we can conclude that computationally and stochastically efficient identification of LPV-SS models on real-world sized problems remains still an open question. Hence, the goal of this paper is to provide a maximum likelihood identification scheme for LPV-SS models in the global, open-loop identification setting, which can provide an integrated solution for both problems. Specifically, to solve Problem 2, we propose to identify surrogate LPV \textit{finite impulse response} (FIR) models via a novel computationally efficient  \textit{correlation analysis} (CRA) method or via an empirical MIMO Bayesian estimation technique. Then, realization of these models is accomplished via a novel basis reduced LPV Ho-Kalman scheme, which grows linearly in complexity compared to previous methods with exponential growth, which are introduced originally in~\cite{Cox2015,Cox2016a}. Next, Problem 1 is solved by integrating the proposed pre-estimation methods into the GB and EM schemes to obtain an ML estimate. In addition, to improve the numerical properties of the GB method, we extend the enhanced Gauss-Newton method~\cite{Wills2008} to the LPV setting. Combining these methods results in a novel three-step approach with a modular structure, achieving both favorable computational properties and enabling ML estimation.

This paper is organized as follows: first, LPV-SS models with general noise structure are analyzed and compared with models relaying on an innovation structure to highlight modeling limitations of the latter form considered in many LPV SID methods. Then, the considered LPV-SS identification problem is introduced (Sec.~\ref{sec:state-space}).
Next, we present our modular identification method, defined in three steps: 
\begin{enumerate*}[label=\arabic*)] 
	\item estimate the FIR model of the underlying system using CRA or MIMO Bayesian estimation (Sec.~\ref{sec:LPV-IIREst}), then
	\item compute an LPV-SS realization based on the estimated coefficients by using a Ho-Kalman like method (Sec.~\ref{sec:RR}), and
	\item to have an ML estimate, refine the LPV-SS model by GB and/or EM optimization (Sec.~\ref{sec:MLref}).
\end{enumerate*}
The contribution of this paper is to provide a detailed overview of the methods applied and to demonstrate that LPV identification of moderate sized models is possible with the proposed scheme. The efficiency of the combined approach is demonstrated by a Monte Carlo study and it is compared to existing LPV-SS identification schemes~\cite{Wingerden2009a,Santos2008,Bianchi2009} (Sec.~\ref{sec:simulation}).

\vspace{-3mm}
\section{The LPV identification problem} \label{sec:state-space}
\vspace{-3mm}

\subsection{Technical preliminaries} \vspace{-2mm}

We define a \textit{random variable} $\mathbf f$ as a measurable function $\mathbf f:\Omega\rightarrow\mathbb{R}^n$, which induces a probability measure $\mathbf{P}$ on $(\mathbb{R}^n,\mathscr{B}(\mathbb{R}^n))$ with an associated Borel measurable space $\mathscr{B}(\mathbb{R}^n)$~\cite{Bogachev2007}. As such, a realization $\nu\in\Omega$ of $\mathbf P$, denoted $\nu\sim\mathbf P$, defines a realization $f$ of $\mathbf f$, i.e., $f:=\mathbf f(\nu) $. A \textit{stochastic process} $\mathbf x$ is a collection of random variables $\mathbf x_t:\Omega\rightarrow\mathbb{R}^n$ indexed by the set $t\in\mathbb{Z}$ (discrete time), given as $\mathbf x=\{ \mathbf x_t : t\in\mathbb{Z} \}$. A realization $\nu_{t}\in\Omega$ of the stochastic process defines a signal trajectory $x:=\{ \mathbf x_t(\nu_{t}) : t\in\mathbb{Z} \}$. We call a stochastic process $\mathbf x$ \textit{stationary} if the probability distribution of $\mathbf x_t$ and joint probability distribution of $(\mathbf x_{t},...,\mathbf x_{t+k})$ for any $k\in\mathbb{N}_+$ are independent of the time-index $t$. In addition, a stationary process consisting of uncorrelated random variables with zero mean and finite variance is called a \textit{white noise process}.
The ring of all real meromorphic functions with finite dimensional domain is denoted by $\mathscr{R}$ and the operator $\diamond:(\mathscr{R},\sP^\sZ)\!\rightarrow\!\mathbb{R}^\sZ$ denotes $(h\diamond p)_t\!=\!h(p_{t+\tau_1},\ldots,p_t,\ldots,p_{t-\tau_2})$ with $\tau_1,\tau_2\!\in\!\mathbb{N}_0$.
The time-shift operator is denoted by $q$, i.e., $qx(t)\!=\!x(t\!+\!1)$, and the set $\{ s,s+1,\cdots,v\}\subset\mathbb{N}_0$ is denoted as $\sI_s^v$.
%

\subsection{The data-generating system} \label{subsec:Model}
\vspace{-2mm}

Consider a \textit{multiple-input multiple-output} (MIMO), discrete-time linear parameter-varying data-generating system, defined by the following first-order difference equation, i.e., LPV-SS representation with general noise model: \vspace{-2mm}
\begin{subequations} \label{eq:SSrep}
\begin{alignat}{3}
  x_{t+1} &= \Afnc(p_t)&x_t&+\Bfnc(p_t)&&u_t+\Gfnc(p_t)w_t, \label{eq:SSrepState} \\
  y_t &= \Cfnc(p_t)&x_t&+\Dfnc(p_t)&&u_t+\Hfnc(p_t)v_t, \label{eq:SSrepOut}
\end{alignat}
\end{subequations}
\vskip -6mm \noindent where $x:\sZ\rightarrow\sX=\Rx$ is the state variable, $y:\sZ\rightarrow\sY=\Ry$ is the measured output signal, $u:\sZ\rightarrow\sU=\Ru$ denotes the input signal, $p:\sZ\rightarrow\sP \subseteq\Rp$ is the scheduling variable, subscript $t\in\mathbb{Z}$ is the discrete time,$w:\sZ\rightarrow\Rx$, $v:\sZ\rightarrow\Ry$ are the sample path realizations of the zero-mean stationary processes: \vspace{-3mm}
\begin{equation} \label{eq:noisy}
\left[ \begin{array}{c}
\mathbf w_t \\ \mathbf v_t
\end{array} \right] \sim \mathcal{N}(0,\Sigma),\hspace{1cm} \Sigma = \left[ \begin{array}{cc} \Qv & \Sv \\ \Sv^\top & \Rv \end{array} \right],
\end{equation} \vskip -7mm \noindent
where $\mathbf w_t:\Omega\rightarrow\Rx$, $\mathbf v_t:\Omega\rightarrow\Ry$ are white noise process, $\Qv\in\mathbb{R}^{\NX\times\NX}$, $\Sv\in\mathbb{R}^{\NX\times\NY}$, and $\Rv\in\mathbb{R}^{\NY\times\NY}$ are covariance matrices, such that $\Sigma$ is positive definite. Furthermore, we will assume $u,p,y,w,v$ to have left compact support to avoid technicalities with initial conditions. As often considered in LPV control theory, the matrix functions $\Afnc(\cdot),...,\Hfnc(\cdot)$, defining the SS representation \eqref{eq:SSrep} are defined as affine combinations: \vspace{-3mm}
\begin{equation}\label{eq:sysMatrices}
\begin{aligned} 
 \Afnc(p_t)&\!=\!A_0+\hspace{-1mm}\sum_{i=1}^{\NPSI} A_i\psi^\ind{i}(p_t),\hspace{-1mm}&\Bfnc(p_t)&\!=\!B_0+\hspace{-1mm}\sum_{i=1}^{\NPSI} B_i\psi^\ind{i}(p_t) , \\
 \Cfnc(p_t)&\!=\!C_0+\hspace{-1mm}\sum_{i=1}^{\NPSI} C_i\psi^\ind{i}(p_t),\hspace{-1mm}&\Dfnc(p_t)&\!=\!D_0+\hspace{-1mm}\sum_{i=1}^{\NPSI} D_i\psi^\ind{i}(p_t), \\
 \Gfnc(p_t)&\!=\!G_0+\hspace{-1mm}\sum_{i=1}^{\NPSI} G_i\psi^\ind{i}(p_t),\hspace{-1mm}&\Hfnc(p_t)&\!=\!H_0+\hspace{-1mm}\sum_{i=1}^{\NPSI} H_i\psi^\ind{i}(p_t), \\[-2mm]
\end{aligned}
\end{equation}
where $\psi^\ind{i}(\cdot):\sP\rightarrow\mathbb{R}$ are bounded scalar functions on $\sP$ and $\{A_i,B_i,C_i,D_i,G_i,H_i\}_{i=0}^{\NPSI}$  are constant, real matrices with appropriate dimensions. Additionally, for well-posedness, it is assumed that $\{\psi^\ind{i}\}_{i=1}^{\NPSI}$ are  linearly independent over an appropriate function space and are normalized w.r.t. an appropriate norm or inner product~\cite{Toth2012}. Due to the freedom to consider arbitrary functions $\psi^\ind{i}$, \eqref{eq:sysMatrices} can capture a wide class of static nonlinearities and time-varying behaviors.

\vspace{-2mm}
\subsection{Properties of LPV-SS representations}
\vspace{-2mm}

In this section, we present some formal definitions needed for the analysis of various noise structures and required definitions for the LPV-IO to LPV-SS realization problem. 

Note that, the deterministic part of~\eqref{eq:SSrep} is governed by \vspace{-4mm}
\begin{subequations} \label{eq:SSrepDet}
\begin{alignat}{3}
  x_{t+1}^\mathrm{d} &= \Afnc(p_t)&x_t^\mathrm{d}&+\Bfnc(p_t)&&u_t, \label{eq:SSrepDetState} \\
  y_t^\mathrm{d} &= \Cfnc(p_t)&x_t^\mathrm{d}&+\Dfnc(p_t)&&u_t. \label{eq:SSrepDetOut}
\end{alignat}
\end{subequations}\vskip -6mm \noindent
The IO solution set, i.e., the \emph{manifest behavior}, of~\eqref{eq:SSrepDet} is \vspace{-3mm}
\begin{multline} \label{eq:SSBehDet}
\mathfrak{B}_{\mathrm{d}}\! =\! \big\{ (y^\mathrm{d},u,p)\!\in\!(\sY\!\times\!\sU\!\times\!\sP)^\mathbb{Z} ~\big|~ \\ \exists x^\mathrm{d}\!\in\!(\sX)^\mathbb{Z} \mbox{ s.t.~\eqref{eq:SSrepDet} holds} \big\}.
\end{multline} \vskip -6mm \noindent
The behavior w.r.t. an LPV-SS representation $\sys$ is denoted $\mathfrak{B}_{\mathrm{d}}(\sys)$. 
On the other hand, the stochastic part of~\eqref{eq:SSrep} is \vspace{-2mm}
\begin{subequations} \label{eq:SSrepStoch}
\begin{alignat}{3}
  \mathbf x_{t+1}^\mathrm{s} &= \Afnc(p_t)&\mathbf x_t^\mathrm{s}&+\Gfnc(p_t)&&\mathbf w_t, \label{eq:SSrepStochState} \\
  \mathbf y_t^\mathrm{s} &= \Cfnc(p_t)&\mathbf x_t^\mathrm{s}&+\Hfnc(p_t)&&\mathbf v_t. \label{eq:SSrepStochOut}
\end{alignat}
\end{subequations}\vskip -6mm \noindent
The manifest behavior corresponding to~\eqref{eq:SSrep} is \vspace{-4mm} 
\begin{multline} \label{eq:SSBeh}
\mathfrak{B}_{\mathrm{SS}}\! =\! \big\{ (y,u,p)\!\in\!(\sY\!\times\!\sU\!\times\!\sP)^\mathbb{Z} ~\big|~  \exists y^\mathrm{d}\!\in\!(\sY)^\mathbb{Z} \mbox{ and} \\[0.5mm] 
\mbox{$\exists\mathbf y^\mathrm{s}_t$ satisfying~\eqref{eq:SSrepStoch} }\mbox{s.t. } (y^\mathrm{d}, u,p)\!\in\!\mathfrak{B}_{\mathrm{d}}   \\[0.5mm] 
\mbox{ and $\forall t$, } \exists \nu\!\in\!\Omega \mbox{ for which } y_t=y^\mathrm{d}_t+\mathbf y^\mathrm{s}_t(\nu)  \big\}.
\end{multline} \vskip -6mm \noindent 
To introduce the essential details of the deterministic realization step in Sec.~\ref{sec:RR}, we momentarily neglect the stochastic process~\eqref{eq:SSrepStoch}. To this end, we take the expectation $\expct\{\mathbf y_t\} = y_t^\mathrm{d}$, which is equivalent as taking $v_t=w_t=0$ in~\eqref{eq:SSrep}. 
In this paper, we are interested in finding an LPV-SS representation with behavoir $\mathfrak{B}_{\mathrm{d}}$ and a minimal state dimension:
\vspace{-2mm}
\begin{defn}[Minimal LPV-SS representations] \label{dfn:minimality} The LPV-SS representation $\sys$~\eqref{eq:SSrep} is called \emph{state minimal}, if there exists no other LPV-SS representation $\sys'$ with $\NX'<\NX$ and equivalent manifest behavior $\mathfrak{B}_{\mathrm{d}}(\sys)=\mathfrak{B}_{\mathrm{d}}(\sys')$. \hfill $\square$
\end{defn} \vskip -2mm \noindent
For specific subclasses of LPV-SS representations in which the functional dependency structure of $\Afnc(\cdot),\ldots,\Dfnc(\cdot)$ is restricted, the minimal state dimension might differ~\cite{Toth2010a}, e.g, when comparing static, affine w.r.t. dynamic, rational dependency (e.g., see \cite[Example 4.1]{Cox2018PHD}). Hence, for the remainder of the paper, state minimality is considered w.r.t. the static, affine dependency in~\eqref{eq:sysMatrices}. 

\vspace{-2mm}
\begin{lem}[Equivalent LPV-SS representations~\cite{Petreczky2017}] \label{lem:SSisomorphism} 
Given two state minimal LPV-SS representations~\eqref{eq:SSrepDet} $\sys$ and $\sys'$ with static, affine dependency~\eqref{eq:sysMatrices} and equivalent state dimensions $\NX=\NX'$. The two representations $\sys$ and $\sys'$ are called \emph{equivalent}, i.e., their associated manifest behaviors are equal $\mathfrak{B}_{\mathrm{d}}(\sys)=\mathfrak{B}_{\mathrm{d}}(\sys')$, if and only if there exists a non-singular isomorphism $T\in\mathbb{R}^{\NX\times\NX}$, such that \vspace{-3mm}
\begin{equation*}
A_i'T=TA_i, \hspace{0.5cm} B_i'=TB_i, \hspace{0.5cm}C_i'T=C_i,\hspace{0.5cm}D_i'=D_i,
\end{equation*} \vskip -6mm \noindent
for all $i\in\sI_0^{\NPSI}$. \hfill $\square$
\end{lem} \vskip -2mm \noindent
Lem.~\ref{lem:SSisomorphism} is a special case of equivalence relation in the LPV case as the transformation matrix $T$ is independent of the scheduling signal, see~\cite[Def. 3.29]{Toth2010a} for the general case. Under dependency structure~\eqref{eq:sysMatrices} and assumption of state minimality, the equivalence class of LPV-SS representations is completely characterized by the non-singular transformation matrix $T$, as given in Lem.~\ref{lem:SSisomorphism}.

We are interested in identification under open-loop conditions, hence, the underlying data-generating system is considered to be asymptotically stable:

\vspace{-2mm}
\begin{defn}[Global asymptotic stability] \label{dfn:assStabil}
An LPV system, represented in terms of~\eqref{eq:SSrepDet}, is called \emph{globally asymptotically stable}, if for all trajectories of $\{u_t,p_t,y^{\mathrm{d}}_t\}$ satisfying~\eqref{eq:SSrepDet}, with $u_t\equiv0$ for $t\geq0$, and $p_t\in\sP$, it holds that $\lim_{t\rightarrow\infty}\vert y_t^{\mathrm{d}}\vert = 0$. \hfill $\square$
\end{defn}

\vspace{-2mm}
\subsection{LPV-SS noise models and the innovation form}
\vspace{-2mm}

A popular model for many subspace identification schemes is the innovation form, e.g., see~\cite{Verhaegen2007}. Under some mild conditions, the LPV-SS representation~\eqref{eq:SSrep} has the following equivalent innovation form:

\vspace{-2mm}
\begin{lem}[LPV-SS innovation form~\cite{Cox2018PHD}] \label{lem:inn} The LPV data-generating system~\eqref{eq:SSrep} can be equivalently represented by a $p$-dependent innovation form \vspace{-2.5mm}
\begin{subequations} \label{eq:SSrepInn}
\begin{alignat}{3}
  \check x_{t+1} &= \Afnc(p_t)&\check x_t&+\Bfnc(p_t)&&u_t+\Kfnc_t\xi_t, \label{eq:SSrepInnState} \\
  y_t &= \Cfnc(p_t)&\check x_t&+\Dfnc(p_t)&&u_t+\xi_t, \label{eq:SSrepInnOut}
\end{alignat}  \vskip -6mm \noindent
where $\xi_t$ is the sample path of $\boldsymbol\xi_t\sim\mathcal{N}(0,\Xi_t)$ and $\Kfnc_t$ can be uniquely determined by \vspace{-3mm}
\begin{align}
\Kfnc_t\! &=\! \left[ \Afnc(p_t) \Pfnc_{t\vert t-1} \Cfnc^\top\!\!(p_t) + \Gfnc(p_t)\Sv\Hfnc^\top\!\!(p_t)\right] \Xi_t^{-1}\!\!, \label{eq-lem:kalmanFilter} \\
\Pfnc_{t+1\vert t}\! &=\! \Afnc(p_t) \Pfnc_{t\vert t-1} \Afnc^\top\!\!(p_t)  - \Kfnc_t \Xi_t \Kfnc^\top_t +\nonumber \\ & \quad\quad \Gfnc(p_t) \Qv \Gfnc^\top\!\!(p_t), \label{eq-lem:innErrorCov} \\
\Xi_t\! &=\! \Cfnc(p_t)\Pfnc_{t\vert t-1}\Cfnc^\top\!\!(p_t) + \Hfnc(p_t)\Rv\Hfnc^\top\!\!(p_t), \label{eq-lem:noiseCov}
\end{align} \vskip -6mm \noindent
\end{subequations}
under the assumption that $\exists t_0\in\mathbb{Z}$ such that $x_{t_0}=0$ and $\Xi_t$ is non-singular for all $t\in[t_0,\infty)$. \hfill $\square$
\end{lem} \vskip -2mm \noindent
In~\eqref{eq-lem:kalmanFilter}-\eqref{eq-lem:noiseCov}, the notation of $\Kfnc_t$, $\Pfnc_{t+1\vert t}$, and  $\Xi_t$ is a shorthand for $\Kfnc_t\coloneqq(\Kfnc\diamond p)_t\in\mathscr{R}^{\NX\times\NY}$, $\Pfnc_{t+1\vert t}\coloneqq(\Pfnc_{t+1\vert t}\diamond p)_t\in\mathscr{R}^{\NX\times\NX}$, and  $\Xi_t\coloneqq(\Xi\diamond p)_t\in\mathscr{R}^{\NY\times\NY}$. The subscript notation $_{t+1\vert t}$ denotes that the matrix function at time ${t+1}$ depends on $p_\tau$ with $\tau\in\{t_0,\ldots,t\}$ where $t\geq t_0$.

In~\cite{Cox2018PHD}, it is shown that the setting of~\eqref{eq:SSrep} is not equivalent to the innovation form with only a static, affine matrix function $\Kfnc(p_t)$, similarly parametrized as~\eqref{eq:sysMatrices}. This static, affine structure is commonly used in many LPV SID methods~\cite{Verdult2002b,Felici2007a,Wingerden2009a}. It follows that, to guarantee state minimality of an innovation form based realization of~\eqref{eq:SSrep}, the Kalman gain $\Kfnc_t$ in~\eqref{eq-lem:kalmanFilter} should have rational and dynamic dependency on $p$. However,~\cite{Cox2018PHD} also shows that a static, affine $\Kfnc(p_t)$ can approximate the general setting~\eqref{eq:SSrepInn} if the state dimension is increased. In practice, we often need to restrict parameterization of $\Kfnc$, e.g., to a static, affine parameterization similar to~\eqref{eq:sysMatrices}, to reduce complexity of the estimation method and variance of the model estimates. Hence, despite the possible increase of state order of the equivalent innovation form, the underlying complexity trade-off might be acceptable from a practical point of view.

\vspace{-2mm}
\subsection{Problem statement} \label{subsec:problemStatement}
\vspace{-2mm}

In this paper, we are interested in identifying LPV-SS models~\eqref{eq:SSrep} with dependency structure as in~\eqref{eq:sysMatrices} to capture the process dynamics of the underlying data-generating system~\eqref{eq:SSrep}. Hence, our focus is not on identifying the noise structure $(\Hfnc,\Gfnc)$, but to derive a methodology which can provide consistent estimates of~\eqref{eq:SSrepDet} under the general noise structure of~\eqref{eq:SSrep}. We will also assume that the scalar functions $\{\psi^\ind{i}\}_{i=1}^{\NPSI}$ are known a priori. As a consequence, we are interested in estimating the parameters of~\eqref{eq:sysMatrices}, i.e. \vspace{-4mm}
\begin{equation}
\Lambda_0 = \left[ \begin{array}{cccccc}
A_0&\ldots&A_\NPSI&B_0&\ldots&B_\NPSI \\ C_0&\ldots&C_\NPSI&D_0&\ldots&D_\NPSI
\end{array} \right],
\end{equation} \vskip -7mm \noindent
with $\Lambda_0\in\mathbb{R}^{\NX+\NY\times(\NX+\NU)(1+\NPSI)}$. Based on these, we denote by $\sys(\Lambda_0)$ the original SS representation of the data-generating system $\sys$ with parameters $\Lambda_0$. According to Lem.~\ref{lem:SSisomorphism}, we aim at identifying an isomorphic $\sys(\Lambda)$ w.r.t. $\sys(\Lambda_0)$, due to the non-uniqueness of the SS representation based on the manifest behavior  $\mathfrak{B}_{\mathrm{d}}(\sys(\Lambda_0))$. Hence, any $\Lambda$ in the following set \vspace{-2mm}
\begin{multline} \label{eq:setOfParameters}
\mathcal{Q}_0=\Bigg\{~ \Lambda ~~\Big|~~ \exists T\in\mathbb{R}^{\NX\times\NX} \mbox{ s.t. } \rank(T)=\NX \mbox{ and } \\
\Lambda = \!\left[\begin{array}{cc} T^{-1} & 0 \\ 0 & I_{\NY} \end{array} \right] \Lambda_0 \left[\begin{array}{cc} I_{1+\NPSI}\otimes T & 0 \\ 0 & I_{\NU(1+\NPSI)} \end{array} \right]\! \Bigg\},
\end{multline} \vskip -6mm \noindent
is considered to be a consistent estimate, where $\otimes$ is the Kronecker product. The set $\mathcal{Q}_0$ is also known as the \textit{indistinguishable parameter set}~\cite{Lee1999,Verdult2003}.

Given a data-set $\Dat=\{u_t,p_t,y_t\}_{t=1}^N$ and the basis functions $\{\psi^\ind{i}\}_{i=1}^{\NPSI}$, our objective is to efficiently find, in a stochastic and computational sense, a model estimate in terms of $\hat\Lambda$ of the data-generating system~\eqref{eq:SSrep} and, accordingly, the state dimension $\NX$. In addition, the proposed scheme should be consistent, i.e., $\hat\Lambda\rightarrow\Lambda\in\mathcal{Q}_0$ with probability one as $N\rightarrow\infty$. We will discuss these properties per individual identification step later on. In the remaining part of this paper, it is assumed that the data-generating LPV-SS system~\eqref{eq:SSrep} with dependency structure~\eqref{eq:sysMatrices} is structurally observable and structurally reachable%
\footnote{See~\cite{Petreczky2017} for a detailed discussion on structural observability and structural reachability in the LPV setting.}%
, i.e., the system is minimal, and that the input-scheduling signals are persistently exciting, such that the parameters are uniquely identifiable up to the indistinguishable parameter set. We will not address the identifiability problem nor we provide persistency of excitation conditions for the input and scheduling signals. A preliminary study can be found in~\cite[Chapter 5]{Cox2018PHD}.


\vspace{-2mm}
\section{Identification of LPV impulse response models} \label{sec:LPV-IIREst}
\vspace{-2mm}

In order to realize our objective defined in Sec.~\ref{subsec:problemStatement}, the first step in the proposed three-step scheme is to capture the SS representation~\eqref{eq:SSrep} or~\eqref{eq:SSrepInn} by its surrogate impulse response representation. It turns out that the coefficients associated with this representation can be captured by linear regression methods. In this section, we present two identification schemes to capture the unknown parameters in a computationally efficient manner by: 1) correlation analysis (Sec.~\ref{subsec:CRA}) or 2) Bayesian impulse response estimation (Sec.~\ref{subsec:MIMOFIR}). The identified impulse response coefficients will be used to realize an SS form (Sec.~\ref{sec:RR}).

\subsection{LPV Impulse response representation}
\vspace{-2mm}

The surrogate \textit{infinite impulse response} (IIR) representation is given as:  \vspace{-2mm}
\begin{lem}[Infinite impulse response~\cite{Toth2010a}] 
\label{lmn:IIR}
Any asymptotically stable LPV system according to Def.~\ref{dfn:assStabil} has a convergent series expansion in terms of the pulse-basis $\{q^{-i}\}_{i=0}^\infty$ given by \vspace{-2mm}
	\begin{equation} \label{eq:IIR}
		y_t = \sum_{i=0}^\infty(h_i\diamond p)_tq^{-i}\,u_t +y_t^\mathrm{s},
	\end{equation} \vskip -5mm \noindent
	where $h_i\in \mathscr{R}^{\NY\times\NU}$ are the expansion coefficient functions, i.e., \emph{Markov coefficients}, and $y_t^\mathrm{s}$ is a sample path of~\eqref{eq:SSrepStoch}. \hfill $\square$
\end{lem} \vskip -3mm \noindent
The IIR of an asymptotically stable LPV-SS representation~\eqref{eq:SSrep} reads as \vspace{-2mm}{\allowdisplaybreaks
\begin{multline}  \label{eq:MarkovParam} 
y_t = \underbrace{\Dfnc(p_t)}_{h_0\diamond p_t}u_t + \underbrace{\Cfnc(p_t)\Bfnc(p_{t-1})}_{(h_1\diamond p)_t}u_{t-1}+ \\
 \underbrace{\Cfnc(p_t)\Afnc(p_{t-1})\Bfnc(p_{t-2})}_{(h_2\diamond p)_t}u_{t-2} + \ldots+ \\
	\underbrace{ \Gfnc(p_t) v_t+ \Cfnc(p_t)\Hfnc(p_{t-1})w_{t-1}+ \ldots}_{y_t^\mathrm{s}},
\end{multline}  } \vskip -6mm \noindent
where $h_i$ converges to the zero function as $i\rightarrow\infty$. The noise $y_t^\mathrm{s}$ in~\eqref{eq:IIR}-\eqref{eq:MarkovParam} is colored, as it is a combination of the IIR filtered innovation noise $w$ and the additive output noise $v$ of~\eqref{eq:SSrep}. Note that the process $y_t^\mathrm{s}$ is quasi-stationary due to the stability of the filters acting on $v_t$ and $w_t$~\cite[Lem. 4.2]{Cox2018PHD}. For notional ease, define $\psi^\ind{i}(p_t)=\psi^\ind{i}_t$, $\psi^\ind{0}_t \equiv1$, and the signal vector $\psi_t=[\begin{array}{cccc} 1&\psi^\ind{1}_t&\cdots&\psi^\ind{\NPSI}_t \end{array}]^\top\in \mathbb{R}^{\NPSI}$. The Markov coefficients can be written as \vspace{-2mm}
\begin{multline}
(h_m\diamond p)_t =\Cfnc(p_t)\Afnc(p_{t-1})\cdots \Afnc(p_{t-m+1})\Bfnc(p_{t-m}) = \\ \sum_{i=0}^\NPSI\!\sum_{j=0}^\NPSI\!\cdots\!\sum_{k=0}^\NPSI\!\sum_{l=0}^\NPSI\! C_iA_j\!\cdots\!A_kB_l\psi^\ind{i}_t 
\psi^\ind{j}_{t-1}\!\ldots\!\psi^\ind{l}_{t-m}, \label{eq:subMarkovParam} 
\end{multline} \vskip -6mm \noindent
where the individual products $C_iA_j\cdots A_kB_l$ are the so-called sub-Markov parameters for $m=1,2,\ldots$. The latter notation is used to denote the effect of the time-shift operator in a product form. The Markov coefficients in~\eqref{eq:MarkovParam} are independent of the parametrization of the matrix functions and the particular state bases, while the sub-Markov parameters are dependent on the parametrization of the functional dependencies in~\eqref{eq:sysMatrices}.


\vspace{-2mm}
\subsection{Correlation analysis} \label{subsec:CRA}
\vspace{-2mm}

The sub-Markov parameters~\eqref{eq:subMarkovParam} can be estimated by \textit{correlation analysis} (CRA), solving the first step of the proposed identification scheme. CRA results in an estimation procedure which grows linearly in the number of data points and is used to estimate each parameter individually. Hence, the correlation based estimation method has a low computational load. CRA makes use of the stochastic property of $u,p,w,v$, hence, in this section, $u$ and $p$ are assumed to be sample paths of stochastic processes $\mathbf u$, $\mathbf p$, respectively. Note that, in such case, $x$ and $y$ obtained from \eqref{eq:SSrep} are sample paths of stochastic processes $\mathbf{x},\mathbf{y}$ which satisfy $\mathbf{x}_{t+1} = \Afnc(\mathbf{p}_t)\mathbf{x}_t+\Bfnc(\mathbf{p}_t)\mathbf{u}_t+\Gfnc(\mathbf{p}_t)\mathbf{w}_t$, and $\mathbf{y}_t = \Cfnc(\mathbf{p}_t)\mathbf{x}_t+\Dfnc(\mathbf{p}_t)\mathbf{u}_t+\Hfnc(\mathbf{p}_t)\mathbf{v}_t$. Furthermore, we introduce $\boldsymbol\psi^\ind{i}_t=\psi^\ind{i}(\mathbf p_t)$, $\boldsymbol\psi^\ind{0}_t \equiv1$, and $\boldsymbol \psi_t = [\begin{array}{cccc} 1 &\psi^\ind{1}(\mathbf p_t) & \cdots & \psi^\ind{\NPSI}_t (\mathbf p_t) \end{array}]^\top$. 
The first step in the CRA is to define the $k$-dimensional cross-correlation. \vspace{-2mm}
\begin{defn}
\label{dfn:NdimCross}
The $k$-dimensional cross-correlation function for the jointly stationary signals $(\mathbf u,\mathbf y,\boldsymbol\psi)$ is defined as \vspace{-2mm}
	\begin{multline*}
	\corr_{y\psi^\ind{s_1},\cdots,\psi^\ind{s_n}u}(\tau_{s_1},\ldots,\tau_{s_n},\tau_u)=  \\
	\expct \left\{ \mathbf y_t \boldsymbol\psi^\ind{s_1}_{t-\tau_{s_1}} \cdots \boldsymbol\psi^\ind{s_n}_{t-\tau_{s_n}}  \left(\mathbf u_{t-\tau_u}\right)^\top\right\},
	\end{multline*}	 \vskip -6mm \noindent
	where $s_i$ is a specific index sequence with $s_1,\ldots,s_n\in\sI_0^\NPSI$ and $\tau_{s_i}\in\sZ_0^+$ is the time-shift associated with the specific basis index $s_i$. \hfill $\square$
\end{defn}
Note that the $k$-dimensional cross-correlation is independent of $t$ due to the assumed joint stationarity of the signals.
\begin{thm}
\label{thm:subMarkov}
The sub-Markov parameters satisfy \vspace{-2mm}
\begin{multline}
C_{s_1}A_{s_2}A_{s_3}\cdots A_{s_{n-1}}B_{s_n} =  \\
\frac{\corr_{y\psi^\ind{s_1},\cdots,\psi^\ind{s_n}u}(0,\ldots,n-1,n-1)}{\sigma^2_{\psi_{s_1}}\cdots~\sigma^2_{\psi_{s_n}}} \Sigma^{-2}_u, \label{eq:CRA-Markov}
\end{multline} \vskip -6mm \noindent
where $\var(\mathbf u)=\Sigma^2_u$, $\sigma^2_{\psi_0}=1$, $\var(\boldsymbol\psi^\ind{i})=\sigma^2_{\psi_i}$, and
\begin{equation} \label{eq:CRA-D}
D_{s_1} = \frac{\corr_{y\psi^\ind{s_1}u}(0,0)}{\sigma^2_{\psi_{s_1}}} \Sigma^{-2}_u,
\end{equation}
where $s_1,\ldots,s_n\in\sI_0^\NPSI$ are the specific index sequences, if the following assumptions hold:\vspace{-2mm}
\begin{enumerate}[label=\bfseries C\arabic*,ref=C\arabic*]
	\item The output signal is generated by a stable LPV system~\eqref{eq:SSrep} with dependency structure~\eqref{eq:sysMatrices}.
	\item\label{ass:whitenoise} The noise processes $\mathbf w$, $\mathbf v$ are distributed as in~\eqref{eq:noisy}.
	\item\label{ass:whiteinput} The input process $\mathbf u$ is a white noise process with finite variance ($\var(\mathbf u)=\Sigma^2_u$) and is independent of $\mathbf w$, $\mathbf v$.
	\item\label{ass:basisfunc} Each process $\boldsymbol\psi^\ind{i}_t \triangleq \psi^\ind{i}_t(\mathbf{p}_t)$ is assumed to be a white noise process with finite variance ($\sigma^2_{\psi_0}=1$, $\var(\boldsymbol\psi^\ind{i})=\sigma^2_{\psi_i}$ for $i=\sI_1^\NPSI$). The processes $\boldsymbol\psi^\ind{i}$ are mutually independent and $\boldsymbol\psi^\ind{i}$ is independent of $\mathbf u$, $\mathbf w$, and $\mathbf v$.
\end{enumerate} \vskip -5mm \noindent \hfill $\square$
\end{thm} \vskip -7mm \noindent
\begin{pf}
See Appendix~\ref{subsec:markovParameter}. \hfill $\blacksquare$
\end{pf}\vskip -4mm \noindent 
Condition \ref{ass:basisfunc} is not over restrictive, e.g., if each $\psi^{\ind{i}}$ is a function of $p^{\ind{i}}$ only, the analytic function $\psi^{\ind{i}}$ is odd and bounded with $\psi^{\ind{i}}(0)=0$, and it is driven by independent white noise scheduling signals $p^{\ind{i}}$ with finite variance, then~\ref{ass:basisfunc} is satisfied.
Note that the sub-Markov parameters in Thm.~\ref{thm:subMarkov} do not depend on the time instant $t$.

The individual sub-Markov coefficients in~\eqref{eq:CRA-Markov} and~\eqref{eq:CRA-D} are estimated by approximating the cross-correlation and variances in Thm.~\ref{thm:subMarkov} based on a finite measured data-set $\mathcal{D}_N$. The variance of the involved signals is estimated by the unbiased sample variance and the $k$-dimensional cross-correlation is approximated via \vspace{-3mm}
	\begin{multline}
	\hat{\corr}_{y\psi^\ind{i},\cdots,\psi^\ind{j}u}(\tau_i,\ldots,\tau_j,\tau_u)=  \\
	\frac{1}{N-\tau_u+1}\sum_{t=\tau_u+1}^N y \psi^\ind{i}_{t-\tau_i}   \cdots \psi^\ind{j}_{t-\tau_j} \left(u_{t-\tau_u}\right)^\top. \label{eq:N-corrEst}
	\end{multline} \vskip -6mm \noindent
It is assumed that the stochastic processes $\mathbf u,\boldsymbol\psi,\mathbf x,\mathbf y,\mathbf w$ are such that $\lim_{N\rightarrow\infty}\hat{\corr}_{y\psi^\ind{i},\cdots,\psi^\ind{j}u}(\cdot)\!=\!\corr_{y\psi^\ind{i},\cdots,\psi^\ind{j}u}(\cdot)$. For example, this assumption holds with probability 1 if $\mathbf u,\boldsymbol \psi,\mathbf x,\mathbf y$ are jointly ergodic. Joint ergodicity has been proven in case $\boldsymbol \psi$ is a random binary  noise and $\mathbf u$ is white noise~\cite{Petreczky2011}.

The proposed CRA method may need a large data-set and $N\gg \tau_u$ such that variance of~\eqref{eq:N-corrEst} is low enough for an accurate parameter estimate. If the process $\mathbf y_t^\mathrm{s}$ in~\eqref{eq:IIR} is a zero mean colored noise, e.g., under the general noise conditions of~\eqref{eq:SSrepStoch}, the CRA estimation is known to be inefficient in the LTI setting~\cite{Ljung1999}, i.e., the variance of the estimated parameters does not correspond to the Cram\'{e}r-Rao bound. The here derived extension to the LPV setting shows that similar statement holds. Therefore, a larger data-set is required to achieve equivalent parameter estimation variance compared to the case when $\mathbf y^\mathrm{s}$ is a white noise with Gaussian distribution. However, an attractive feature of the method is that the sub-Markov parameters can be estimated individually and the computational complexity scales with $\mathcal{O}\left( N(2+n_\mathrm{y}^2+\NY\NU n) \right)$ where $n$ is the amount of specific index sequences $\{s_1,\ldots,s_n\}$ \footnote{Unbiased sample variance scales with $\mathcal{O}(2N+Nn_\mathrm{y}^2)$ and~\eqref{eq:N-corrEst} scales with $\mathcal{O}(N\NY\NU n)$.} . Hence, the problem scales linearly in $N$, $\NU$, $n$ and quadratic in $\NY$. We will see that for the basis reduced Ho-Kalman method only a subset of the sub-Markov parameters are needed for realization. Hence, the combination of the LPV-SS realization scheme with the CRA significantly reduces the computational demand, as identification of the full impulse response is omitted.

\vspace{-2mm}
\subsection{Bayesian impulse response estimation} \label{subsec:MIMOFIR}
\vspace{-2mm}

As an alternative to CRA, the sub-Markov parameters can be estimated using a Tikhonov regression based LPV-FIR estimation procedure, where the optimal regularization matrix is determined in a Bayesian way with a Gaussian prior, i.e., \textit{Bayesian LPV-FIR estimation}. In addition, the Bayesian framework allows to estimate the functional dependencies $\psi^\ind{i}(\cdot)$ in a nonparamtric way~\cite{Golabi2014,Darwish2015}. However, for the sake of simplicity, we consider that these functions are known a priori.

\vspace{-2mm}
\subsubsection{The truncated IIR model}
\vspace{-2mm}

In the Bayesian framework, Eq.~\eqref{eq:IIR} is approximated by the following finite order truncation: \vspace{-2mm}
\begin{equation} \label{eq:FIRmodel}
	y_t \approx \sum_{i=0}^\NH(h_i\diamond p)_tu_{t-i} +y_t^\mathrm{s},
\end{equation} \vskip -6mm \noindent
with $\NH>0$. Eq.~\eqref{eq:FIRmodel} corresponds to a \textit{finite impulse response} (FIR) model of~\eqref{eq:IIR} with order $\NH$. Due to the convergence of $h_i$, approximation error of~\eqref{eq:FIRmodel} can be chosen arbitrary small by selecting $\NH$. Furthermore, define \vspace{-2mm}
\begin{equation} \label{eq:ReachM}
  \M_1 \!=\!\left[\hspace{-0.5mm}\begin{array}{ccc} B_0&\hspace{-0.5mm}\mbox{\small{\dots}}&\hspace{-1mm}B_\NPSI\end{array} \hspace{-1mm}  \right]\!\!,~~ \M_j \!=\!\left[\hspace{-0.5mm}\begin{array}{ccc} A_0\M_{j-1}&\hspace{-1mm}\mbox{\small{\dots}}&\hspace{-1mm}A_\NPSI\M_{j-1}  \end{array}\hspace{-1 mm}\right]\!\!. 
 \end{equation} \vskip -6mm \noindent
Based on \eqref{eq:subMarkovParam} and \eqref{eq:FIRmodel}, the samples in $\Dat$ satisfy the following relationship: \vspace{-2mm}
\begin{equation} \label{eq:FIRmatrix}
\Ym_N  = \thetam_0~\Phim_N +\Vm_N,
\end{equation} \vskip -8mm \noindent
with \vspace{-2mm}{\allowdisplaybreaks
\begin{align*} 
\Ym_{\!\!N} \!\!&=\!\! \left[ \hspace{-1mm}\begin{array}{ccc} y_{\NH+1}&\hspace{-0.5mm}\mbox{\small{\dots}} &\hspace{-0.5mm} y_N \end{array} \hspace{-0.5mm}\right]\!\!,  \hspace{1cm}\Vm_{\!N}= \left[\hspace{-0.5mm} \begin{array}{ccc} y_{\NH+1}^\mathrm{s}&\hspace{-0.5mm}\mbox{\small{\dots}} &\hspace{-0.5mm} y_{M}^\mathrm{s} \end{array} \hspace{-0.5mm}\right]\!\!, \\ 
\thetam_0 \!\!&=\!\!\left[\hspace{-0.5mm} \begin{array}{ccccccccc} D_0 &\hspace{-0.5mm}\mbox{\small{\dots}} &\hspace{-0.5mm} D_\NPSI & C_0\M_1 & \hspace{-0.5mm}\mbox{\small{\dots}} & \hspace{-0.5mm}C_\NPSI\M_1 & C_0\M_2 & \hspace{-0.5mm}\mbox{\small{\dots}} &\hspace{-0.5mm}  C_\NPSI\M_\NH \end{array} \hspace{-0.5mm}\right]\!\!, \\ 
\Phim_{\!N} \!\! &=\!\! \left[\hspace{-1mm}\begin{array}{ccc}
\psi_{\NH+1}\!\otimes \!u_{\NH+1} &\hspace{-2mm}\mbox{\small{\dots}} &\hspace{-2mm} \psi_{\!N}\!\otimes\! u_{\!N} \\
\psi_{\NH+1}\!\otimes\!\psi_\NH\!\otimes\! u_{\NH} & \hspace{-2mm}\mbox{\small{\dots}} &\hspace{-2mm} \psi_{\!N}\!\otimes\!\psi_{\!N-1\!}\!\otimes\! u_{\!N-1\!} \\
\vdots & \hspace{-2mm}\small{\ddots} & \hspace{-2mm}\vdots \\
\psi_{\NH+1}\!\otimes\!\mbox{\small{\dots}}\!\otimes\!\psi_{1}\!\otimes\! u_{1} &\hspace{-0.5mm} \mbox{\small{\dots}} & \hspace{-0.5mm}\psi_{M}\!\otimes\!\mbox{\small{\dots}}\!\otimes\!\psi_{N-\NH}\!\otimes\! u_{N-\NH} 
\end{array}\hspace{-1mm}\right]\!\!, 
\end{align*}} \vskip -6mm \noindent
where $M=N-\NH-1$, $\Ym_N\in\mathbb{R}^{\NY\times M}$ are the measured outputs, $\thetam_0\in\mathbb{R}^{\NY\times\NF}$ is the collection of the to-be-estimated sub-Markov parameters with $\NF = \sum_{i=1}^{\NH+1}(1+\NPSI)^i\NU$, $\Phim_N\in\mathbb{R}^{\NF\times M}$ is the regression matrix and $\Vm_N\in\mathbb{R}^{\NY\times M}$ is the stacked noise realization. The resulting output predictor of the MIMO FIR model~\eqref{eq:FIRmatrix} can be written as \vspace{-2mm}
\begin{equation} \label{eq:FIRvector}
\hat Y_N  = \Phi^\top_N\theta,
\end{equation} \vskip -6mm \noindent
where $\NTH = \NY\NF$, $\hat Y_N\in\mathbb{R}^{\NY M \times 1}$ is the predicted output, $\Phi^\top_N=\Phim^\top_N\otimes\eye_{\NY}\in\mathbb{R}^{\NY M\times\NTH}$,  and $\theta\in\mathbb{R}^{\NTH \times 1}$. For notational reasons, also introduce $Y_N \hspace{-0.5mm}=\hspace{-0.5mm} \vecM(\Ym_N)$, $\theta_0 \hspace{-0.5mm}=\hspace{-0.5mm} \vecM(\thetam_0)$, and $W_N \hspace{-0.5mm}=\hspace{-0.5mm} \vecM(\Vm_N)$.

\vspace{-2mm}
\subsubsection{Tikhonov regression based estimate}
\vspace{-2mm}

Even in the LTI case, a well-known issue in estimation of FIR models via the least-squares approach is the high variance of the estimated parameters, due to the relatively large number of parameters required to adequately represent the process dynamics. $\ell_2$ regularization makes it possible to control the so-called \emph{bias-variance trade-off}, i.e., dramatically decrease the variance by introducing a relatively small bias on the estimates~\cite{Ljung2013}. The corresponding \emph{weighted Ridge regression} or Tikhonov regularization problem is given by \vspace{-2mm}
\begin{equation}
\min_\theta \Vert \Phi^\top_N\theta -Y_N \Vert^2_{W_\mathrm{e}} + \Vert\theta\Vert^2_{W_\mathrm{r}}, \label{eq:minimization}
\end{equation} \vskip -7mm \noindent
where $\Vert x\Vert_W=  \sqrt{x^\top W x}$ denotes the \emph{weighted Euclidean norm}, hence, the first term in~\eqref{eq:minimization} corresponds to a weighted $\ell_2$ norm of the prediction-error of~\eqref{eq:FIRmodel}, while the second term is the weighted $\ell_2$ norm of $\theta$. Both $W_\mathrm{e},W_\mathrm{r}\in\mathbb{R}^{\NTH\times\NTH}$ are positive semi-definite (symmetric) regularization matrices and the analytic solution of \eqref{eq:minimization} is \vspace{-2mm}
\begin{equation} \label{eq:thetRWLS}
\hat{\theta}_\RWLS = \left(\Phi_N W_\mathrm{e} \Phi^\top_N +W_\mathrm{r} \right)^{-1}\Phi_N W_\mathrm{e} Y_N.
\end{equation}\vskip -6mm \noindent
The regularization matrix $W_\mathrm{r}$ is chosen such that $\Phi_N W_\mathrm{e} \Phi^\top_N +W_\mathrm{r}$ is invertible. If $W_\mathrm{r}=0$, $W_\mathrm{e}=I$, and $\mathbf y^\mathrm{s}$ is a white noise process with Gaussian distribution then~\eqref{eq:thetRWLS} is the least squares solution, which results in the asymptotically efficient, unbiased, ML estimate.

Analogous to CRA, if $W_\mathrm{r}=0$, $W_\mathrm{e}=I$, and the additive noise $\mathbf y^\mathrm{s}$ is a zero mean coloured noise process, but uncorrelated with the input and scheduling signals; then the estimator is unbiased, although it is inefficient in terms of increased variance. If $\mathbf y^\mathrm{s}$ and $\mathbf u$ are correlated, then an LPV \textit{instrumental variable} (IV) estimator can be used to remove the bias, e.g., see~\cite{Laurain2010}. 

\vspace{-2mm}
\subsubsection{A Bayesian way of optimizing regularization}
\vspace{-2mm}

One of the main questions with the application of regularization is how to choose the regularization matrix $W_\mathrm{r}$, such that an optimal bias-variance trade-off is found. A recently introduced efficient data-driven approach follows an \textit{empirical Bayes method}~\cite{Carlin1996}. Let us assume in this section that the process noise is zero, i.e., $\mathbf w=0$ in~\eqref{eq:SSrep}. Hence, the output additive noise process $\mathbf v$ in~\eqref{eq:SSrep} is equal to the output additive noise $\mathbf y^\mathrm{s}$ in~\eqref{eq:IIR} and~\eqref{eq:FIRmodel} corresponds to an output error setting. Furthermore, assume that the parameter vector $\theta_0$ is a random variable with \emph{Gaussian distribution}:\vspace{-3mm}
\begin{equation*}
\boldsymbol\theta_0\sim\mathcal{N}(\theta_\mathrm{a},P_\alpha), \hspace{0.5cm} \theta_\mathrm{a}=0,
\end{equation*} \vskip -6mm \noindent
where the covariance matrix $P_\alpha$ is a function of some hyper parameters $\alpha\in \mathbb{R}_+^{n_\alpha}$. In the Bayesian setting, under the assumption that $u$ and $p$ are given realizations, $\Phi_N$ is deterministic and, according to~\eqref{eq:FIRmatrix}, the output vector $Y_N$ and the parameters $\theta_0$ are jointly Gaussian variables: \vspace{-2mm}
\begin{equation} \label{eq:jointGaussianVariables}
\left[\hspace{-1mm} \begin{array}{c} \boldsymbol\theta_0 \\ \mathbf Y_{\!N} \end{array} \hspace{-1mm}\right]\hspace{-1mm}\sim\hspace{-0.5mm}\mathcal{N}\hspace{-0.5mm}\left( \hspace{-0.5mm}\left[\hspace{-1mm}\begin{array}{c} 0 \\ 0 \end{array}\hspace{-1mm}\right]\hspace{-1mm},\hspace{-1mm} \left[\hspace{-1mm}\begin{array}{cc} P_{\alpha} & P_{\alpha} \Phi_N \\ \Phi_N^\top P_{\alpha} &\hspace{3mm} \Phi_N^\top P_{\alpha} \Phi_N \hspace{-0.5mm}+\hspace{-0.5mm} \eye_M\hspace{-0.5mm}\otimes\hspace{-0.2mm}\Rv \end{array}\hspace{-1mm}\right]\hspace{-0.5mm}\right)\!\!,\hspace{-1mm}
\end{equation} \vskip -6mm \noindent
with $\Rv$ as in~\eqref{eq:noisy}.
It can be shown that the maximum \textit{posteriori} estimate and the minimal variance estimate of $\boldsymbol\theta_0$ given $\mathbf Y_{\!N}$ is equivalent to the weighted regularized least squares estimate $\hat{\theta}_\RWLS$~\eqref{eq:thetRWLS}, e.g., see~\cite{Ljung2013}, if the weighting and regularization matrices are chosen as \vspace{-2mm}
\begin{equation} \label{eq:weighting2Bays}
W_\mathrm{e} = \eye_M\otimes\Rv^{-1},\hspace{1cm} W_\mathrm{r} = P_{\alpha}^{-1}.
\end{equation} \vskip -6mm \noindent
This connection makes it possible to create an estimate of $\Rv$ and $P_{\alpha}$ from data that minimizes the marginal likelihood w.r.t.~\eqref{eq:jointGaussianVariables}. Notice that covariance matrix $P_{\alpha}$, parametrized by $\alpha$, and the noise covariance matrix $\Rv$ satisfy \vspace{-2mm}
\begin{equation}
\mathbf Y_{\!N} \sim \mathcal{N}\left(  0, \Phi_N^\top P_{\alpha} \Phi_N + \eye_M\otimes\Rv \right).
\end{equation} \vskip -6mm \noindent
Hence, the likelihood function of the observation $Y_N$ given $\alpha$ and  $\Rv$ can be used to arrive to their posteriori estimate: \vspace{-2mm}
\begin{multline}
\hat{\alpha} = \argmax_{\alpha} f(Y_N\vert\alpha) = \argmin_{\alpha} -2\log f(Y_N\vert\alpha) \\=  \argmin_{\alpha} \log\left(\det\left( \Phi_N^\top P_{\alpha} \Phi_N + \eye_M\otimes\Rv \right)\right) \\[-1mm] + Y_N^\top \left( \Phi_N^\top P_{\alpha} \Phi_N + \eye_M\otimes\Rv\right)^{-1}Y_N, \label{eq:loglikelihood}
\end{multline}\vskip -6mm \noindent
where the constant terms are excluded and $f(\cdot)$ is the probability density function of the multivariate normal distribution. For a detailed description of pros and cons of the empirical Bayes method compared to other methods, see~\cite{Pillonetto2014}.

The choice of the parametrization of $P_{\alpha}$ is of big importance as it governs the ``quality'' of the estimate. The matrix $P_\alpha=\theta_0^\top\theta_0$ will give the lowest parameter mean-squared-error (MSE)\footnote{The parameter mean-squared-error (MSE) for an estimator is defined as $\MSE(\hat\theta_N)=\expct\{(\hat\theta_N-\theta_0)(\hat\theta_N-\theta_0)^\top\}$.}~\cite{Eldar2006}. However, the true system parameters $\theta_0$ are unknown. Therefore, $P_{\alpha}$ is often chosen to be a parameterized kernel function to characterize an appropriate search space for an optimal choice of $P_{\alpha}$. Many different kernel functions can be employed for this purpose, see~\cite{Chen2012} for a detailed discussion. For the sake of simplicity, in this paper, we aim at Ridge regression, i.e., we will use $P_{\alpha}=\alpha \eye$. Regularized regression, in general, is know to provide estimates with a lower parameter MSE compared to non-regularized methods, like the CRA method. On the other hand, for the regularized regression, the complete model needs to be estimated, from which, as we will see later, not all parameters are necessary for realization. Consequently, the combination of regularized regression with LPV-SS realization loses computational efficiency compared to the CRA method with LPV-SS realization, but it is applicable under a much wider set of conditions (e.g., we can relax~\ref{ass:whitenoise}-\ref{ass:basisfunc} in Thm.~\ref{thm:subMarkov}).

\vspace{-2mm}
\section{A basis reduced Ho-Kalman SS realization}\label{sec:RR}
\vspace{-2mm}

The aforementioned identification schemes of Sec.~\ref{subsec:CRA} and~\ref{subsec:MIMOFIR} can consistently estimate the sub-Markov paramters of~\eqref{eq:SSrep} under mild assumptions. However, to achieve our goal; an efficient LPV-SS realization of the estimated FIR model is needed. In~\cite{Toth2012}, the well-known Ho-Kalman realization scheme is extended to the LPV case for realizing LPV-SS models with static and affine dependence on the scheduling variable. However, the size of the $l$-step extended observability and $k$-step extended reachability matrices grow exponentially in $l,k$ and grow polynomially in the scheduling dimension $\NPSI$. Recently, we proposed a basis reduced Ho-Kalman scheme~\cite{Cox2015}, where only the non-repetitive parts of the extended Hankel matrix are selected, which drastically decreases the computational load, compared to the full realization scheme of~\cite{Ho1966,Toth2012}. The resulting scheme does not depend on any approximations, hence, it is an exact, deterministic realization scheme, and will be briefly explained in this section.

Given a set of sub-Markov parameters associated with the deterministic part~\eqref{eq:SSrepDet}. To indicate which sub-Markov parameters of the involved extended reachability, observability, and Hankel matrices are selected, we introduce a string of characters, called a \emph{selection}, to denote the considered matrices and their order of multiplication. To define the set of considered strings of characters, introduce $[\sI_s^v]^n$ as the set of all $n$-length sequences of the form $(i_1,\ldots,i_n)$ with $i_1,\ldots,i_n\in\sI_s^v$. The elements of $\sI_s^v$ will be viewed as characters and the finite sequences of elements of $\sI_s^v$ will be referred to as strings. Then $[\sI_s^v]^n$ is the set of all strings containing exactly $n$ characters. The string $\alpha\in\left[\sI_0^\NPSI\right]^n_0$ is called a selection with $n\geq0$ where $\left[\sI_0^\NPSI\right]^n_0=\{\epsilon\} \cup \sI_0^\NPSI\cup\ldots\cup\left[\sI_0^\NPSI\right]^n$ and $\epsilon$ denotes the empty string. Define by $\numstr{\alpha}$ the amount of characters of a single string. Applying a sequence $\alpha$ will give the ordering of multiplication of matrices $\{A_i\}_{i=0}^\NPSI$: if $\alpha=\epsilon$, then $A_\epsilon=I$ else
 \vspace{-2mm}
\begin{equation} \label{eq:prodMat}
A_\alpha = \prod_{i=1}^{\numstr{\alpha}}A_{\left[\alpha\right]_i}=A_{\left[\alpha\right]_1}A_{\left[\alpha\right]_2} \cdots A_{\left[\alpha\right]_{\numstr{\alpha}}},
\end{equation} \vskip -6mm \noindent
where $\left[\alpha\right]_i$ denotes the $i$-th character of the string $\alpha$. As an example, let us define the set $\left[\sI_0^1\right]^2_0=\{\epsilon\} \cup \sI_0^\NPSI\cup\left[\sI_0^\NPSI\right]^2=\{\epsilon,0,1,00,01,10,11\}$. Take, for instance, $\alpha = 10\in\left[\sI_0^1\right]^2_0$ which indicates $A_\alpha =A_1A_0$.
Based on this selection, the $(i,j)$-th element of a single sub-Markov parameter $C_\gamma A_\alpha B_\beta\in\mathbb{R}^{\NY\times\NU}$ is denoted by $C_\gamma^\ind{i} A_\alpha B_\beta^\ind{j}$ for $\alpha\in[\sI_0^\NPSI]_0^n$, $\beta\in\sI_0^\NPSI$, $\gamma\in\sI_0^\NPSI$. Then a selection of the extended reachability matrix is represented by \vspace{-2mm}
\begin{align}
\selR=\left\{(\alpha_1,\beta_1,j_1),\ldots,(\alpha_\NR,\beta_\NR,j_\NR)\right\},
\end{align} \vskip -7mm \noindent
where $\alpha_1,\ldots,\alpha_\NR\in\left[\sI_0^\NPSI\right]^n_0$, $\beta_1,\ldots,\beta_\NR\in\sI_0^\NPSI$, and $j_1,\ldots,j_\NR\in\sI_1^\NU$. The length of the string $\alpha_i$ may vary. Using this basis, a sub-matrix of the extended reachability matrix~\cite{Verdult2002a} is selected, defined by \vspace{-3mm}
\begin{equation}
\reach_\selR = \left[ \begin{array}{cccc} A_{\alpha_1}B_{\beta_1}^\ind{j_1} & \hspace{2mm} A_{\alpha_2}B_{\beta_2}^\ind{j_2} & \hspace{2mm}\ldots & \hspace{2mm} A_{\alpha_\NR}B_{\beta_\NR}^\ind{j_\NR} \end{array} \right],
\end{equation} \vskip -7mm \noindent
where $\reach_\selR\in\mathbb{R}^{\NX\times\NR}$ and $\ind{j_k}$ denotes the $j_k$-th column of $B_{\beta_k}$ for $k=1,\ldots,\NR$. 
Analogously, a basis of the extended observability matrix is selected by \vspace{-2mm}
\begin{equation}
\selO=\left\{(i_1,\gamma_1,\alpha_1),\ldots,(i_\NO,\gamma_\NO,\alpha_\NO)\right\},
\end{equation}\vskip -7mm \noindent
where $\alpha_1,\ldots, \alpha_\NO\in\left[\sI_0^\NPSI\right]^n_0$, $\gamma_1,\ldots,\gamma_\NO\in\sI_0^\NPSI$, and $i_1,\ldots,i_\NO\in\sI_1^\NY$. Note that $\alpha_i$ in $\selR$ and $\selO$ can be different. This selection $\selO$ defines the sub-matrix of the extend observability matrix as \vspace{-3mm}
\begin{equation}
\obsv_\selO = \left[ \begin{array}{ccc} \left(C_{\gamma_1}^\ind{i_1}A_{\alpha_1}\right)^\top &\hspace{2mm}\cdots & \hspace{2mm} \left(C_{\gamma_\NO}^\ind{j_\NO}A_{\alpha_\NO}\right)^\top \end{array} \right]^\top,
\end{equation} \vskip -7mm \noindent
where $\obsv_\selO\in\mathbb{R}^{\NO\times\NX}$ and $\ind{i_k}$ denotes the $i_k$-th row of $C_{\gamma_k}$ for $k=1,\ldots,\NO$.
The sets $\selR$ and $\selO$ are chosen appropriately, such that $\rank(\reach_\selR)=\NX$, $\rank(\obsv_\selO)=\NX$, and hence $\rank(\obsv_\selO\reach_\selR)=\NX$. If this condition is satisfied, then we call the selection $\selR$ and $\selO$ a \textit{basis selection}. For such a basis selection $(\selO,\selR)$, define \vspace{-2mm}
\begin{equation}
\begin{aligned} \label{eq:HankBas}
\hank_{\selO,\selR} &= \obsv_\selO\reach_\selR, &\hspace{0.5cm}\hank_{\selO,\selR,k} &= \obsv_\selO A_k\reach_\selR,  \\
\hank_{\selO,k} &= \obsv_\selO B_k, &\hank_{k,\selR} &= C_k\reach_\selR,
\end{aligned}
\end{equation} \vskip -6mm \noindent
where $\hank_{\selO,\selR}\in\mathbb{R}^{\NO\times\NR}$, $\hank_{\selO,\selR,k}\in\mathbb{R}^{\NO\times\NR}$, $\hank_{\selO,k}\in\mathbb{R}^{\NO\times\NU}$ and $\hank_{k,\selR}\in\mathbb{R}^{\NY\times\NR}$. Note that these sub-Hankel matrices in~\eqref{eq:HankBas} are composed of the sub-Markov parameters. \vspace{-2mm}

\begin{lem} \label{lem:basisSSReal}
Define a column selection $\selR$ with $\NR=\NX$ and a row selection $\selO$ with $\NO\geq\NX$ such that $\rank(\hank_{\selO,\selR})=\NX$. The set of matrices \vspace{-2mm}
\begin{equation} \label{eq-lem:basisSSReal}
\begin{aligned}
\hat{A}_k &= \hank^\dagger_{\selO,\selR}\hank_{\selO,\selR,k}, &\hspace{0.5cm}\hat{B}_k = \hank^\dagger_{\selO,\selR}\hank_{\selO,k}, \\
\hat{C}_k &= \hank_{k,\selR},
\end{aligned}
\end{equation} \vskip -6mm \noindent
for $k\in\sI_0^\NPSI$ give a joint minimal LPV-SS representation of $\sys$ in~\eqref{eq:SSrep} with the dependency structure~\eqref{eq:sysMatrices}, i.e., \vspace{-4mm}
\begin{equation} \label{eq-lem:realMatrices}
\left[ \begin{array}{cccccc}
\hat A_0&\ldots&\hat A_\NPSI&\hat B_0&\ldots&\hat B_\NPSI \\ \hat C_0&\ldots&\hat C_\NPSI&D_0&\ldots&D_\NPSI
\end{array} \right] \in\mathcal{Q}_0.
\end{equation}  \vskip -6mm \noindent
In~\eqref{eq-lem:basisSSReal}, $\hank^\dagger_{\selO,\selR}$ denotes the left pseudo inverse of $\hank_{\selO,\selR}$. \hfill $\square$
\end{lem} \vskip -6mm \noindent
\begin{pf}
As $\hank^\dagger_{\selO,\selR}$ exists and $\hank_{\selO,\selR}$ has full column rank, the proof is straightforward by applying the isomorphism $T=\reach^{-1}_\selR$. \hfill $\blacksquare$
\end{pf} \vskip -4mm \noindent
From the practical and numerical point of view, a reliable implementation of~\eqref{eq-lem:basisSSReal} follows by using \textit{singular value decomposition} (SVD). Define a basis selection $\NR,\NO\geq\NX$ with $\rank(\hank_{\selO,\selR})=\NX$ and compute an economical SVD: $\hank_{\selO,\selR}=U_\NX\Sigma_\NX V^\top_\NX$. Then a realization of $\sys$ is \vspace{-2mm}
\begin{equation}\label{eq:RealBasisSVD}
\begin{aligned}
\hat{A}_k &= \hat{\obsv}^\dagger_\selO\hank_{\selO,\selR,k}\hat{\reach}^\dagger_\selR, 
&\hat{B}_k &= \hat{\obsv}^\dagger_\selO\hank_{\selO,k}, \\
\hat{C}_k &= \hank_{k,\selR}\hat{\reach}^\dagger_\selR, &&
\end{aligned}
\end{equation} \vskip -7mm \noindent
for $k\!\in\!\sI_0^\NPSI$ with pseudo inverses $\hat{\reach}^\dagger_\selR=V_\NX\Sigma_\NX^{-1/2}$, $\hat{\obsv}^\dagger_\selO=\Sigma_\NX^{-1/2}U_\NX^\top$. Realization~\eqref{eq:RealBasisSVD} gives an LPV-SS representation of $\sys$ in~\eqref{eq:SSrep}, i.e., $\{\hat{A}_0,\ldots,\hat{C}_\NPSI\}$ satisfies~\eqref{eq-lem:realMatrices}. The proof of this methodology can be found in~\cite{Cox2015}.

In case the sub-Hankel matrices~\eqref{eq:HankBas} are filled with estimated sub-Markov parameters, the state order $\NX$ can be chosen based upon the magnitude of the singular values $\Sigma_\NX$, i.e., an approximate realization (e.g., see~\cite{Kung1978}). Note that the realization in~\eqref{eq:RealBasisSVD} does not have any restrictions on the maximum amount of columns chosen $\NR\geq\NX$, compared to Lem.~\ref{lem:basisSSReal} where $\NR=\NX$. Hence, the rank-revealing property of the SVD of $\hank_{\selO,\selR}$ allows to find a reliable estimate of $\NX$.

This bases reduced realization can considerably decrease the size of the Hankel matrix and, therefore, reduce the computational load, compared to the realization with the full Hankel matrix~\cite[Eq. (48)]{Toth2012}. In the basis reduced realization, the SVD is only applied on a $\NO\times\NR$ matrix instead of a matrix with size $\NY\sum_{l=1}^i(1+\NPSI)^l\times\NU\sum_{l=1}^j(1+\NPSI)^l$ in the full realization case. Note that $\NO,\NR=\NX$ in the ideal case, which gives the computational lower bound that is similar to the LTI case. The amount of sub-Markov parameters in~\eqref{eq:HankBas} is $\NO\NR+(1+\NPSI)(\NO\NR+\NO\NU+\NY\NR)$, which increases linearly in all parameters $\NPSI,\NR,\NO,\NU,\NY$, compared to $\NY\sum_{l=1}^i(1+\NPSI)^l\cdot\NU\sum_{l=1}^j(1+\NPSI)^l$, which grows exponentially with increasing $i$ and $j$ and polynomially with increasing $\NPSI$. To illustrate, the realization of a system with input/output dimension $\NY=\NU=2$, state dimension $\NX=4$, and scheduling dimension $\NPSI=5$, the full Hankel matrix $\hank_{2,2}$ has 7056 elements, while the sub-Hankel matrices for $\NR=\NO=10$ have only 940 elements.

\vspace{-2mm}
\section{Maximum likelihood refinement} \label{sec:MLref}
\vspace{-2mm}

The basis reduced Ho-Kalman realization cannot guarantee that the LPV-SS model realized from the identified sub-Markov parameters is a maximum likelihood estimate, even if the underlying approaches are capable of providing ML estimates. Hence, to reach the maximum likelihood LPV-SS model estimate, two solutions are explored for refinement:
\begin{enumerate*}[label=\arabic*)]
	\item the gradient-based (GB) search method, and
	\item the expectation maximization (EM) algorithm.
\end{enumerate*}
Both methods are nonlinear iterative optimization techniques and cannot be used as stand alone methods, as they are prone to local minima. For example,~\cite[Table III]{Wills2008} shows the number of failed model identification iterations for inefficient initial estimates in an LTI-SS identification problem. Hence, Step 1 and Step 2 of our proposed identification scheme, i.e., LPV impulse response estimation with LPV-SS realization, can be seen as a numerically efficient method for initializing GB or EM methods. The efficiency of this combination will be shown in Sec.~\ref{sec:simulation}.


\vspace{-2mm}
\subsection{Gradient based PEM}
\vspace{-2mm}

PEM methods aim at minimizing the mean-squared prediction-error criterion w.r.t. the free model parameters. For LPV-SS models, the minimization problem is nonconvex and nonunique based upon $\Dat$ \cite{Cox2018PHD}. The optimization is usually solved via a gradient-based search strategy such as a Newton or similar type of method. In this paper, the enhanced Gauss-Newton based search method of~\cite{Wills2008} is extended to the LPV case. The enhanced Gauss-Newton includes: 
\begin{enumerate*}[label=\arabic*)]
	\item an automated strategy of regularization and SVD truncation on the Jacobian matrix to obtain a search direction,
	\item an Armijo line search backtracking rule, and 
	\item lowering the dimension of the parameter space by using the \textit{data-driven local coordinate} (DDLC) frame.
\end{enumerate*}
The DDLC frame is the ortho-complement of an affine approximation of the indistinguishable set $\mathcal{Q}_0$ around the current model parameters. Consequently, the DDLC ensures that the nonlinear optimization does not wander among parameterizations of SS models with equivalent manifest behavior. Additionally, the DDLC results in a minimal parametrization in the LTI case and, hence, the PEM optimization problem is of minimal dimension~\cite{Cox2018PHD}.  The combination of improved gradient-based search strategies and the DDLC frame increases the computational demand per iteration, however, in general, it significantly improves the convergence rate.

\vspace{-2mm}
\subsection{Expectation Maximization}
\vspace{-2mm}

The key element of the EM method is to presume the existence of a complete data-set $Z_N=(Y_N,X_N)$, which contains not only the actual observations $Y_N$, but also the missing state-sequence $X_N$. The iterative EM method identifies LPV-SS models by considering the state-sequence as the missing data. With this choice, the maximization of the ML is a joint estimation problem and is solved in an alternating manner. EM methods for the LTI case have been developed in~\cite{Shumway1982,Watson1983,Gibson2005b} and an LPV extension of EM is given in~\cite{Wills2011}. We can apply~\cite{Wills2011} under the assumption that the noise structure in the data-generating system~\eqref{eq:SSrep} is with $\Hfnc(p)=I$ and $\Gfnc(p)=I$. We provide here a brief overview of the main steps of this algorithm. Each iteration of the EM consist of two steps:
\begin{enumerate*}[label=\arabic*)]
	\item the expectation, and
	\item the maximization step.
\end{enumerate*} 
In the expectation step, given the current model estimate, the likelihood of the complete data-set conditional on the data observed is approximated. The likelihood, i.e., obtaining the unknown state trajectory $x_t$, can be estimated via various approaches, e.g., particle filtering~\cite{Schon2011}, or Kalman filtering~\cite{Gibson2005,Shumway2010}. In the example section, we provide the comparison using an implementation with the Kalman filter, the Kalman smoother, and a one-lag covariance smoother similar to~\cite{Wills2011}.
In the second step (maximization step), the approximated likelihood is maximized with respect to the model parameters. As the state-sequence is known, the estimation problem becomes linear-in-the-parameters with an analytic solution.
The EM method is relatively straightforward to implement and the computational load scales linearly with the data-set length. The EM algorithm usually converges rapidly in early stages, but its rate of convergence near the maximum is substantially lower than of the GB method, e.g., see~\cite{Watson1983,Gibson2005}.


\vspace{-2mm}
\section{Simulation Example} \label{sec:simulation}
\vspace{-2mm}

In this section, the performance of the proposed three-step identification procedure is assessed on a Monte-Carlo simulation study using a randomly generated stable LPV-SS model in innovation form with scheduling independent matrix function, i.e., $\Kfnc(p_t)=K$. The Monte-Carlo study shows the performance of the methods in the following cases: \vspace{-2mm}
\begin{enumerate}[leftmargin=6mm]
	\item Correlation analysis with basis reduced Ho-Kalman LPV-SS realization (without refinement step),
	\item Correlation analysis with basis reduced Ho-Kalman LPV-SS realization and EM or GB refinement step,
	\item Bayesian FIR estimation with basis reduced Ho-Kalman LPV-SS realization (without refinement step),
	\item Bayesian FIR estimation with basis reduced Ho-Kalman LPV-SS realization and EM or GB refinement step.
\end{enumerate} \vskip -2mm \noindent
The proposed procedure is compared to state-of-the-art LPV-SS identification methods, such as the predictor-based subspace identification (PB)~\cite{Wingerden2009a}, successive approximation identification algorithm (SA)~\cite{Santos2008}, and the robust identification/invalidation method (RI)~\cite{Bianchi2009}. Furthermore, the estimated SS model by these approaches is refined, identical to the case of CRA and FIR, by using the estimated SS model as initialization for the EM or GB method. This shows which approach can provide better initialization for the ML step and how far the delivered models are from the ML estimate. The case study is performed on a Macbook pro laptop, late 2013 with an 2.6GHz Intel i5 processor and Matlab 2014b. For the comparison, the scripts provided by the authors of~\cite{Santos2008,Wingerden2009a,Bianchi2009} are used.

\vspace{-2mm}
\subsection{Data-generating system and model structure}
\vspace{-2mm}

The data-generating system is randomly selected in terms of an SS model~\eqref{eq:SSrepInnState}-\eqref{eq:SSrepInnOut} with input-output dimensions $\NU=\NY=2$, scheduling dimension $\NPSI=5$, minimal state dimension $\NX=4$, and  affine dependence, i.e., the known basis functions are $\psi^\ind{i}=p^\ind{i}$ with $p^\ind{i}$ denoting the $i^\mathrm{th}$ element of $p$. The SS model represented system has a scheduling independent innovation matrix, i.e., $\Kfnc(p_t)=K$. This simplified innovation form is chosen, for the sake of fairness of the comparison, as all aforementioned methodologies are able to consistently identify this particular representation, except the EM methodology due to its different noise assumptions. The system was constructed such that~\eqref{eq:SSrep} and the innovation form based output substituted equation \vspace{-2mm}
\begin{equation*}
\check x_{t+1} = (\Afnc(p_t)\!-\!K\Cfnc(p_t))\check x_t\!+\!(\Bfnc(p_t)\!-\!K\Dfnc(p_t))u_t\!+\!Ky_t,
\end{equation*} \vskip -6mm \noindent
are asymptotically input-to-state stable on the domain $p_t\in\mathbb{P}=[-1, 1]^5$ with a quadratic Lyapunov function defined by a constant symmetric matrix~\cite{Scherer1996}. The LPV-SS model is available at~\cite{Cox2018DataSet}.

\vspace{-2mm}
\subsection{Identification setting}
\vspace{-2mm}

The identification data-set is generated with a white $\mathbf u$ with uniform distribution $\mathbf u_t\sim\uniform(-1,1)$, and white $\mathbf p$ with random binary distribution on $(-0.9,0.9)$, each of length $N=5\cdot10^3$. The noise process $\boldsymbol\xi$ is taken as a white noise with distribution $\boldsymbol\xi_t\sim\mathcal{N}(0,\Vv)$ where $\Vv$ is diagonal and it is chosen such that the \emph{signal-to-noise ratio} (SNR) \vspace{-2mm}
\begin{equation*}
	\SNR_y^{\ind{i}} = 10 \log \frac{\sum_{t=1}^N (y^{\ind{i}}_t)^2 }{\sum_{t=1}^N (y^{\mathrm{s},\ind{i}}_t)^2},
\end{equation*} \vskip -7mm \noindent
is set for various Monte-Carlo experiments as $\SNR_y^{\ind{i}}=\{40,25,10,0\}$\,dB for all $i=1,\ldots,\NY$. The $\ind{i}$ denotes the $i$-th channel, i.e., element of the vector signal, and $\SNR_y^{\ind{i}}$ is the SNR of the output $y^{\ind{i}}$. In this setting, the signals are jointly ergodic and the parameters can be consistently identified~\cite{Petreczky2011}. The performance of the scheme is tested on a validation data-set $\Dval$ of length $N_{\mathrm{val}}=200$ with different excitation conditions than the estimation data-set: \vspace{-2mm}
\begin{align}
u_t  &\!=\! \left[ \begin{array}{c} 0.5\cos(0.035t) \\  0.5\sin(0.035t) \end{array} \right]+\delta_{t,u}, \label{eq:valDataInput} \\
p^\ind{i}_t &\!=\! 0.25\!-\!0.05i\!+\!0.4\sin\!\left(\!0.035t \!+\! \frac{2i\pi}{5}\!\right)\!\!+\!\delta_{t,p_i}, \label{eq:valDataScheduling}
\end{align} \vskip -6mm \noindent
where $\delta_{t,u}\in\mathbb{R}^\NU$, $\delta_{t,p_i}\in\mathbb{R}$ are element wise i.i.d. sequences with uniform distribution $\uniform(-0.15,0.15)$. To study the statistical properties of the developed identification scheme, a Monte-Carlo study with $N_{\mathrm{MC}}=100$ runs is carried out, where in each run a new realization of the input, scheduling, and noise sequences are taken. In each run, all considered methods are applied on the identification data-set. The data-set is available at~\cite{Cox2018DataSet}. We will asses the performance of the CRA, FIR, and RI model estimates without refinement step by comparing the simulated output $\hat{y}$ of the estimated model to the noise free output $y_t^\mathrm{d}$. In all other cases, the one-step-ahead predicted output $\hat{y}$ of the estimated model is compared to the one-step-ahead predicted output of an oracle predictor (i.e., the one-step-ahead predicted output using the original data-generating system). This dichotomy in assessing different signals is caused by the fact that the CRA, FIR, and RI do not identify a noise model, hence, the one-step-ahead predicted output is equal to the simulated output, therefore, comparing it to the noise free output $y_t^\mathrm{d}$ of the process part is more adequate. On the other hand, the remaining methods include an estimate of a noise model, thus the estimated plant and noise model are assessed by using the one-step-ahead predictor. In this case, the achieved results are compared w.r.t. the oracle, as its generated output is the maximum achievable output estimate given the data-set. The performance criterion used is the \textit{best fit rate} (BFR)\footnote{Usually the BFR is defined per channel. Eq.~\eqref{eq:BFR} is the average fit performance over all channels.}\vspace{-2mm}
\begin{equation}
\BFR=\max\hspace{-1mm}\left\{\hspace{-0.5mm}1\hspace{-1mm}-\hspace{-1mm}\frac{\frac{1}{N}\hspace{-1mm}\sum_{t=1}^N\hspace{-1mm}\Vert y_t-\hat{y}_t\Vert_2}{\frac{1}{N}\hspace{-1mm}\sum_{t=1}^N\hspace{-1mm}\Vert y_t-\bar{y}\Vert_2}, 0\hspace{-0.5mm}\right\} \cdot 100\%,\label{eq:BFR}
\end{equation} \vskip -6mm \noindent
using $\Dval$. In \eqref{eq:BFR}, $\bar{y}$ defines the mean of the predicted/true output $y_t$ in $\Dval$ and $\hat{y}_t$ is the simulated output of the model w.r.t.~\eqref{eq:valDataInput} and~\eqref{eq:valDataScheduling} in $\Dval$. Next, we will provide a summary of the used design parameters, which are optimized to provide the highest $\BFR$. The FIR model order is chosen as $\NH=2$ with $P_\alpha=\alpha I$. The hyperparameter $\alpha$ is tuned by using the Bayesian MIMO formulation of~\cite{Cox2016a}. In the realization step, the basis reduced Ho-Kalman scheme uses $\NO=\NR=10$ bases, where the controllability matrix is spanned by $\selR=\!\{\!(\epsilon,0,2),(\epsilon,1,2),(\epsilon,2,1),(\epsilon,2,2),\ldots,(\epsilon,5,2)\!\}$ and the observability is spanned by $\selO = \{ (1,0,\epsilon),\ldots,(2,1,\epsilon),\\(2,2,\epsilon),(1,3,\epsilon),\ldots,(1,4,\epsilon),(1,5,\epsilon),(2,5,\epsilon)\}$. The basis of the Hankel matrix is selected by using the entries of the full Hankel matrix with the largest absolute value. For the PB method, the future $f$ and past window $p$ are chosen as $f=p=3$. For the SA method, the number of block rows in the Hankel matrix is chosen to be 4 and the iterative procedure is stopped if the 2-norm of the eigenvalues of the $A_0$ matrix do not change more than $10^{-6}$ or if it exceeds 100 iterations. For the RI method, only the first 150 data samples are taken into account as the computational complexity of the problem does not allow to use all data points of $\Dat$. For the EM method, the relative and absolute tolerance on the marginal log likelihood are chosen as $2\cdot10^{-3}$ and $10^4$, respectively, with a maximum of $20$ iterations. For the GB method, we use $\beta=10^{-4}$, $\gamma=0.75$, $\eta_{\mathrm{min}}=10^{-5}$, $\alpha_{\mathrm{min}}=0.001$, $\nu=0.01$, $\epsilon=10^{-6}$ according to the notation of~\cite{Wills2008}, and a maximum of $20$ iterations.

\vspace{-2mm}
\subsection{Analysis of the results} 
\vspace{-2mm}

Table~\ref{tab:results} shows the mean and the standard deviation of the $\BFR$ on $\Dval$ and execution time of the estimation algorithms per Monte Carlo run for different $\SNR_y=\{40,25,10,0\}$\,dB. Similar results are obtained w.r.t. the simulation error, however, due to space limitations it is not presented. Note that the SA method does not often converge to the considered system with $\NP=5$, hence, also a simulation study is done where the system to be identified had only $\NP=2$ scheduling signals (SA2). In addition, remark that, the RI method only identifies $\Cfnc(\cdot),\Dfnc(\cdot)$ and assumes $\Afnc(\cdot),\Bfnc(\cdot)$ to be known. \\
The table shows that the FIR with bases reduced realization outperforms the CRA, PB, SA, and RI methods. The CRA performs worse, because regularized methods, such as FIR, provide estimates with lower parameter MSE by tuning the bias/variance trade-off. However, this tuning comes with an increased computational cost of approximately 4 times. \\
The PB is outperformed by the FIR, as it needs to estimate significantly more parameters, which is a well known problem of this method~\cite[Table 1]{Wingerden2009a}. Estimation of the increased amount of parameters also results in an increased computational load for this method. On the other hand, PB can identify unstable systems as only the one-step-ahead predictor dynamics are required to be stable and it can also be used in a closed-loop identification setting. \\ 
The SA method has, in many cases, problems with convergence. Presumably, this is caused by using an LTI subspace method to initialize the iterative scheme. 
The method has a substantially higher $\BFR$ and less convergence problems if the data-generating system has $\NP=2$ instead of $\NP=5$. \\
The RI method can potentially outperform the other methods, as the $\Afnc(\cdot),\Bfnc(\cdot)$ matrix functions are a-priori known. However, the computational complexity of the RI method only allows to use a small portion of the data-set $\Dat$ for estimation (in our case 150 out of 5000), hence, a large decrease in its performance is seen for lower SNRs. \\
All performance criteria indicate that the additional refinement step, with the EM or GB method, will lead to a better estimate of the model, as expected. Only in case of the $\SNR_y^{\ind{i}} = 0$\,dB noise scenario, the EM refinement step does not improve the estimate. In this case, the EM method is not able to converge due to the large noise contribution. The GB method outperforms the EM method in all cases. Partially, this might be caused by the additional steps to improve the numerical properties of the GB method, i.e., the automated strategy of regularization and SVD truncation of the Jacobian matrix and line search backtracking rule. Furthermore, the underlying data-generating system is not within the noise model set of the EM methodology, leading to a suboptimal filter with a lower achieved $\BFR$ compared to GB. Therefore, no fair conclusions can be drawn on the relative performance of EM w.r.t. GB based on this simulation study.
In addition, we would like to highlight that the CRA and FIR are not statistically efficient under the considered noise scenario, as they do not identify a noise model. 
Hence, it is impressive that these methods are capable of providing efficient initializations of PEM, even under a non-idealistic noise scenario.

Summarizing, the proposed three-step approach results in a maximum-likelihood estimate with a lower computational time and higher performance compared to existing state-of-the-art LPV-SS identification approaches.

\begin{table*}[!t] 
\caption{The mean and the standard deviation (between parentheses) of the $\BFR$ and execution time of the estimation algorithms per Monte Carlo run for different $\SNR_y^{\scriptstyle\ind{i}}=\{40,25,10,0\}$dB is given. The $\BFR$ is based on the one-step-ahead predicted output of the estimated model on the validation data-set except the methods with an asterisk for which it is based on the simulated output.  The correlation analysis (CRA), finite impulse response (FIR) estimation, the predictor-based subspace identification (PB), successive approximation identification (SA), and the robust identification/invalidation (RI) method are used and refinement of the estimates is performed by expectation-maximization (EM) or gradient based (GB) algorithm. The SA2 indicates the results for the SA method where the system to be identified had $\NP=2$ scheduling signals. For this table, $N_{\mathrm{MC}}=100$ Monte Carlo simulations are performed. The number in the superscript indicates how many successful trails have been achieved out of the $100$ runs.}
\begin{center}
	\input{total.tex}
\end{center}\label{tab:results}
\end{table*}


\vspace{-2mm}
\section{Conclusion} \label{sec:conclusion}
\vspace{-2mm}

In this paper, we have presented a computationally efficient, modular three-step LPV-SS identification approach, which contains the following steps: 
\begin{enumerate*}[label=\arabic*)] 
	\item estimation of the Markov coefficient sequence using correlation analysis or a Bayesian FIR estimation, then
	\item efficient LPV-SS realization by using a basis reduced Ho-Kalman method, and 
	\item refinement of the LPV-SS model estimate by a GB or EM optimization methodology.
\end{enumerate*}	
This three-step approach can consistently identify the underlying data-generating system. The effectiveness of the scheme has been demonstrated on a real-world sized MIMO LPV-SS model identification problem under harsh noise conditions and it has been compared to other methods. Any combination of the scheme was able to identify the system within seconds, significantly faster than its competitors while also achieving better performance.

\vspace{-2mm}
\begin{ack}                               
\vspace{-2mm}
We would like to thank the authors of~\cite{Wingerden2009a,Santos2008,Bianchi2009} for providing their code to make the simulation study possible.
\end{ack}
\vspace{-2mm}

\bibliographystyle{ieeetr}        
\bibliography{library.bib}           

\appendix

\vspace{-2mm}
\section{Proof of Theorem~\ref{thm:subMarkov}} \label{subsec:markovParameter}
\vspace{-2mm}

The proof is based on computing the expected value of the cross-correlation between the stationary signals $\mathbf y,\boldsymbol\psi,\mathbf u$ under the assumption that the signals are ergodic. First, the relation for the direct feed-through matrices $D_{s_1}$ is shown. Let us substitute the IIR~\eqref{eq:IIR} for $\mathbf y$ in $\corr_{y\psi^\ind{s_1}u}(0,0)$, which gives \vspace{-3mm}
\begin{align}
&\corr_{y\psi^\ind{s_1}u}(0,0) = \nonumber\\
&\!=\!\expct\left\{\left(\Dfnc(\mathbf p_t)\mathbf u_t \!+\! \Cfnc(\mathbf p_t)\Bfnc(\mathbf p_{t-1})\mathbf u_{t-1}\!+\! \cdots\!+\!\mathbf y^\mathrm{s}_t\right)\hspace{-0.5mm}\boldsymbol\psi^\ind{s_1}_t\mathbf u^\top_t\right\} \nonumber\\
&\!=\!\expct\left\{\left(D_0+\sum_{i=1}^{\NPSI} D_i\boldsymbol\psi^\ind{i}_t \right)\mathbf u_t \boldsymbol\psi^\ind{s_1}_t\mathbf u^\top_t\right\} + \nonumber\\
&\hspace{1mm}\expct \left\{ \Cfnc(\mathbf p_t)\Bfnc(\mathbf p_{t-1})\left(\mathbf u_{t-1}\right)\boldsymbol\psi^\ind{s_1}_t\mathbf u^\top_t \right\} \!+\! \cdots \!+\! \expct\{\mathbf y^\mathrm{s}_t \boldsymbol\psi^\ind{s_1}_t\mathbf u^\top_t\} \nonumber\\
&\!=\!D_{s_1} \sigma^2_{\psi_{s_1}} \Sigma^2_u. \label{eq-app:provesubMarkD}
\end{align} \vskip -6mm
\noindent%
Eq.~\eqref{eq-app:provesubMarkD} holds due to the whiteness property of the processes $(\mathbf u,\boldsymbol\psi)$ and their independence. Also see that $\expct\{\mathbf y^\mathrm{s}_t \boldsymbol\psi^\ind{s_1}_t\mathbf u^\top_t\}=0$, as $\mathbf w$, $\mathbf v$, and $\boldsymbol\psi$ are assumed to be independent of $\mathbf u$ and $\mathbf y^\mathrm{s}$ satisfies the relation given in~\eqref{eq:MarkovParam}, therefore, $\mathbf y^\mathrm{s}$ is independent from $\mathbf u$. Hence, $\expct\{\mathbf y^\mathrm{s}_t \boldsymbol\psi^\ind{s_1}_t\mathbf u^\top_t\}=\expct\{\mathbf y^\mathrm{s}_t \boldsymbol\psi^\ind{s_1}_t\}\expct\{\mathbf u^\top_t\}=0$. For all other sub-Markov parameters, let us consider the following formulation \vspace{-3mm}
\begin{align}
 &\corr_{y\psi^\ind{s_1},\cdots,\psi^\ind{s_n}u}(0,\ldots,n-1,n-1) = \nonumber \\
&\hspace{4mm}=\expct \Big\{\left(\Dfnc(\mathbf p_t)\mathbf u_t + \Cfnc(\mathbf p_t)\Bfnc(\mathbf p_{t-1})\mathbf u_{t-1}+ \cdots+\mathbf y^\mathrm{s}_t\right) \nonumber \\
&\hspace{4mm}\qquad\qquad\qquad \boldsymbol\psi^\ind{s_1}_{t-\tau_{s_1}}\cdots\boldsymbol\psi^\ind{s_n}_{t-\tau_{s_n}}\left(\mathbf u_{t-\tau_u}\right)^\top\Big\} \nonumber \\
&\hspace{4mm}=\expct \Big\{ C_{s_1}A_{s_2}\cdots A_{s_{n-1}}B_{s_n} \left(\boldsymbol\psi^\ind{s_1}_{t-\tau_{s_1}}\right)^2\cdots \nonumber \\
& \hspace{4mm}\qquad\qquad\qquad \left(\boldsymbol\psi^\ind{s_n}_{t-\tau_{s_n}}\right)^2\mathbf u_{t-\tau_u}\left(\mathbf u_{t-\tau_u}\right)^\top \Big\} + \nonumber \\
&\hspace{8.4mm} \expct\! \left\{\!\left(\Dfnc(\mathbf p_t)\mathbf u_t\!+\!\cdots\!+\!\mathbf y^\mathrm{s}_t \right)\! \boldsymbol\psi^\ind{s_1}_{t-\tau_{s_1}}\!\cdots\boldsymbol\psi^\ind{s_n}_{t-\tau_{s_n}}\!\left(\mathbf u_{t-\tau_u}\right)^{\!\top} \right\} \nonumber \\
&\hspace{4mm}= C_{s_1}A_{s_2}\cdots A_{s_{n-1}}B_{s_n} \sigma^2_{\psi_{s_1}}\cdots~\sigma^2_{\psi_{s_n}} \Sigma^{2}_u. \label{eq-app:provesubMark}
 \end{align} \vskip -6mm
Reordering~\eqref{eq-app:provesubMarkD} and~\eqref{eq-app:provesubMark} concludes the proof.

\begin{rem}
It is possible to get the same sub-Markov parameters with different multiplications of $\boldsymbol\psi_{s_i}$ and corresponding shifts, e.g., $\corr_{y\psi^\ind{s_1}\psi^\ind{s_2}u}(0,1,1)$ gives the same sub-Markov parameter $C_{s_1}B_{s_2}$ as $\corr_{y\psi^\ind{s_1},\cdots,\psi^\ind{s_4}u}(0,1,4,4,1)$. In scope of the estimation of these sub-Markov parameters, we impose the above given ordering to keep the multiplications with $\boldsymbol\psi^\ind{s_i}$ minimal. \hfill $\square$
\end{rem}

\end{document}

%% file: total.tex
{ \small
\setlength\tabcolsep{1.3mm} 
\begin{tabular}{|l||ll|ll|ll|ll||ll|ll|ll|} \hline
&\multicolumn{8}{c||}{ $\boldsymbol{\BFR}$ [\%]} & \multicolumn{4}{c|}{\textbf{Time Elapsed} [s]}\\ \cline{2-13}
&\multicolumn{2}{c|}{$40$dB} & \multicolumn{2}{c|}{$25$dB} & \multicolumn{2}{c|}{$10$dB} & \multicolumn{2}{c||}{$0$dB}  & \multicolumn{2}{c|}{$25$dB}  &\multicolumn{2}{c|}{$0$dB}\\ \hline
CRA$^*$   & $81.47$ & ($4.106$)   & $81.44$ & ($4.433$)  & $81.14$ & ($4.504$)  & $75.07$ & ($5.823$) & $2.269$ & ($0.1481$) & $2.229$ & ($0.1519$) \\ 
CRA + EM  & $99.71$ & ($0.05576$) & $98.80$ & ($0.1710$) & $91.31$ & ($0.5744$) & $74.13$ & ($1.804$) & $8.297$ & ($0.2722$) & $8.351$ & ($0.2978$) \\ 
CRA + GB  & $99.86$ & ($0.02888$) & $99.27$ & ($0.1600$) & $95.81$ & ($0.8313$) & $87.81$ & ($2.513$) & $8.626$ & ($0.6105$) & $12.94$ & ($1.146$) \\ \hline
FIR$^*$   & $99.32$ & ($0.1466$)  & $98.74$ & ($0.2872$) & $94.19$ & ($1.346$)  & $83.55$ & ($3.868$) & $10.87$ & ($0.5037$) & $9.098$ & ($0.4474$) \\
FIR + EM  & $99.73$ & ($0.05327$) & $98.80$ & ($0.1679$) & $91.26$ & ($0.5432$) & $74.14$ & ($1.788$) & $16.88$ & ($0.6055$) & $15.21$ & ($0.3635$) \\
FIR + GB  & $99.86$ & ($0.02886$) & $99.27$ & ($0.1600$) & $95.81$ & ($0.8313$) & $87.81$ & ($2.513$) & $17.21$ & ($0.8603$) & $19.62$ & ($1.236$) \\ \hline
PB        & $95.90$ & ($1.185$)   & $86.25$ & ($2.761$)  & $75.89$ & ($3.051$)  & $61.35$ & ($8.831$) & $88.47$ & ($0.5274$) & $88.43$ & ($0.4078$) \\
PB + EM   & $98.50$ & ($0.2035$)  & $98.02$ & ($0.2849$) & $92.39$ & ($1.703$)  & $78.80$ & ($4.875$) & $92.84$ & ($0.5456$) & $92.79$ & ($0.4136$) \\
PB + GB   & $99.79$ & ($0.05832$) & $99.27$ & ($0.1600$) & $95.81$ & ($0.8312$) & $87.57$ & ($2.893$) & $97.62$ & ($1.317$)  & $100.2$ & ($0.4811$) \\ \hline
SA2       & $82.97$ & $\mbox{(}12.90\mbox{)}^{~77}$   & $83.83$ & $\mbox{(}9.771\mbox{)}^{~83}$  & $83.04$ & $\mbox{(}10.70\mbox{)}^{~56}$ & $73.78$ & $\mbox{(}12.04\mbox{)}^{~37}$ & $21.52$ & ($8.165$)  & $26.86$ & ($3.883$) \\
SA2 + GB  & $97.29$ & $\mbox{(}13.56\mbox{)}^{~77}$   & $99.51$ & $\mbox{(}0.1216\mbox{)}^{~81}$ & $96.68$ & $\mbox{(}2.363\mbox{)}^{~53}$  & $90.49$ & $\mbox{(}2.249\mbox{)}^{~37}$ & $26.94$ & ($8.126$)  & $34.21$ & ($3.687$) \\ \hline
SA        & $65.57$ & $\mbox{(}4.225\mbox{)}^{~18}$   & $65.55$ & $\mbox{(}3.283\mbox{)}^{~13}$  & $66.18$ & $\mbox{(}3.630\mbox{)}^{~24}$  & $53.24$ & $\mbox{(}8.806\mbox{)}^{~29}$ & $108.9$ & ($3.442$)  & $107.0$ & ($0.3076$) \\
SA + GB   & $97.75$ & $\mbox{(}8.483\mbox{)}^{~16}$   & $99.31$ & $\mbox{(}0.1306\mbox{)}^{~13}$ & $94.51$ & $\mbox{(}6.373\mbox{)}^{~24}$  & $87.51$ & $\mbox{(}2.358\mbox{)}^{~27}$ & $118.0$ & ($3.615$)  & $122.3$ & ($1.793$) \\ \hline
RI$^*$   & $99.25$ & ($0.1876$)  & $95.75$ & ($0.9813$) & $76.11$ & ($5.991$)  & $26.66$ & ($15.25$) & $103.4$ & ($5.261$)  & $108.9$ & ($6.406$) \\
RI + EM  & $99.73$ & ($0.05359$) & $98.80$ & ($0.1710$) & $91.29$ & ($0.5441$) & $74.34$ & ($1.663$) & $109.3$ & ($5.307$)  & $114.8$ & ($6.438$) \\
RI + GB  & $99.86$ & ($0.02887$) & $99.27$ & ($0.1600$) & $95.81$ & ($0.8313$) & $87.81$ & ($2.513$) & $110.3$ & ($5.352$)  & $121.3$ & ($6.465$) \\ \hline
\end{tabular}
}